\documentclass{aa}  
\usepackage{graphicx}
\usepackage{txfonts}
\usepackage{natbib}
\bibpunct{(}{)}{;}{a}{}{,}
\usepackage[colorlinks=true, citecolor=blue]{hyperref}
\usepackage{multirow}
\makeatletter
\renewcommand*\aa@pageof{, page \thepage{} of \pageref*{LastPage}}
\makeatother
\usepackage{etoolbox}
\makeatletter
\patchcmd\@combinedblfloats{\box\@outputbox}{\unvbox\@outputbox}{}{\errmessage{\noexpand patch failed}}
\makeatother
\usepackage{array,booktabs}
\begin{document} 


\title{Extragalactic megahertz-peaked spectrum radio sources at milliarcsecond scales}
\author{M.~A.~Keim,
          \inst{1, 2}
          J.~R.~Callingham,
          \inst{2}
          \and
          H.~J.~A.~Röttgering
          \inst{1}
          }
\institute{Leiden Observatory, Leiden University, P.O. Box 9513, 2300 RA Leiden, The Netherlands\\
              \email{keim@strw.leidenuniv.nl}
            \and
             ASTRON, Netherlands Institute for Radio Astronomy, Oude Hoogeveensedijk 4, 7991PD, Dwingeloo, The Netherlands\\
             }
 \date{Received 14 June 2019; accepted 16 July 2019}


\abstract{Extragalactic peaked-spectrum radio sources are thought to be the progenitors of larger, radio-loud active galactic nuclei (AGN). Synchrotron self-absorption (SSA) has often been identified as the cause of their spectral peak. The identification of new megahertz-peaked spectrum sources from the GaLactic and Extragalactic All-sky Murchison Widefield Array (GLEAM) survey provides an opportunity to test how radio sources with spectral peaks below 1 GHz fit within this evolutionary picture. We observed six peaked-spectrum sources selected from the GLEAM survey, three that have spectral characteristics which violate SSA and three that have spectral peaks below 230 MHz, with the Very Long Baseline Array at 1.55 and 4.96 GHz. We present milliarcsecond resolution images of each source and constrain their morphology, linear size, luminosity, and magnetic field strength. Of the sources that are resolved by our study, the sources that violate SSA appear to be compact doubles, while the sources with peak frequencies below 230 MHz have core-jet features. We find that all of our sources are smaller than expected from SSA by factors of $\gtrsim$20. We also find that component magnetic field strengths calculated from SSA are likely inaccurate, differing by factors of $\gtrsim$5 from equipartition estimates. The calculated equipartition magnetic field strengths more closely resemble estimates from previously studied gigahertz-peaked spectrum sources. Exploring a model of the interaction between jets and the interstellar medium, we demonstrate that free-free absorption (FFA) can accurately describe the linear sizes and peak frequencies of our sources. Our findings support the theory that there is a fraction of peaked-spectrum sources whose spectral peaks are best modelled by FFA, implying our understanding of the early stages of radio AGN is incomplete.}

\authorrunning{Keim, M.~A. et al.}
\maketitle


\section{Introduction} \label{Sec:Introduction}

The jets of active galactic nuclei (AGN) are energized by the accretion of matter onto supermassive black holes which are thought to exist at the center of all massive galaxies~\citep{1964ApJ...140..796S,2003ApJ...589L..21M}. Despite the ubiquity of AGN in the radio sky, several key questions remain unanswered about their life-cycle including the duration of radio activity, how radio jets are launched by the AGN, and how the released energy impacts the host galaxy~\citep{2015MNRAS.453.1223H, 2018MNRAS.478L..89D, 2017NatAs...1..596M}. In particular, the early evolutionary stages of a radio AGN are the subject of ongoing debate~\citep{2016AN....337....9O,2018MNRAS.477..578C,2018MNRAS.475.3493B}. Compact radio doubles identified through very-long-baseline interferometry (VLBI) have been suggested to be the progenitors of Fanaroff-Riley (FR) type I and type II radio-loud AGN~\citep{1982A&A...106...21P, 1974MNRAS.167P..31F} due to their resemblance at kpc- and pc-scales~\citep{1995A&A...302..317F, 1997AJ....113..148O}. Such compact doubles are associated with the spectral classes referred to as high-frequency peaked (HFP), gigahertz-peaked spectrum (GPS), and compact steep spectrum (CSS) sources. HFP sources have spectral peaks above 5 GHz and pc-scale radio morphologies~\citep{2000A&A...363..887D}. GPS sources have spectral turnovers around 1 GHz and linear sizes ${\lesssim}$1 kpc~\citep{1991ApJ...380...66O}. Finally, CSS sources peak at frequencies below 1 GHz and can extend to ${\sim}$20 kpc~\citep{1990A&A...231..333F}. Nearly 1500 Megahertz-peaked spectrum (MPS) sources with definite spectral peaks observed at frequencies below 1 GHz have also been recently identified~\citep{2017ApJ...836..174C}.

It is suspected that each of these spectral classes essentially describe the same physical sources observed at different evolutionary stages. In the proposed evolutionary model based on linear size, spectral peak, and radio power, HFP sources evolve into GPS sources, which in turn grow into CSS sources and, depending on luminosity, finally become either FR I or FR II galaxies~\citep{2010MNRAS.408.2261K}. This `youth' model is supported by studies of kinematic age based on observed hotspot separation over multiple epochs~\citep{1998A&A...337...69O, 1999NewAR..43..669O, 2003PASA...20...69P, 2005ApJ...622..136G} and radiative age estimated from magnetic field strength and break frequency~\citep{1999A&A...345..769M,2010MNRAS.402.1892O} which have found ages ranging from ${\sim}10^{1}$ to ${\sim}10^{5}$ years. 

However, several lines of evidence suggest that the `youth' model of peaked-spectrum sources is incomplete. For example, population studies suggest an excess of peaked-spectrum sources compared to larger radio sources~\citep{1991ApJ...380...66O,2000MNRAS.319..445S,2012ApJ...760...77A}. While such excess could imply extremely short activity periods, an alternative to youth is the frustration model in which the compact size is due to confinement by a dense circum-nuclear medium~\citep{1984AJ.....89....5V}. Observational studies of individual sources have attempted to rule out either model. For instance, X-ray emission from 3C186 implied it is unlikely that the pressure of gas in the surrounding cluster medium was sufficient to confine the radio components~\citep{2005ApJ...632..110S}, while spectral modeling of the GPS source PKS B0008-421 found that it was likely surrounded by a dense medium~\citep{2015ApJ...809..168C}. It is also possible that both scenarios could apply to sources in the population~\citep{2017ApJ...836..174C}.

Intrinsic to the debate over the `youth' vs. `frustration' scenarios is the absorption mechanism responsible for the observed spectral turnover. Typically synchrotron self-absorption (SSA) via the relativistic electrons themselves will cause a source to become optically thick at low frequencies with a characteristic spectral index $\alpha$ limit of $2.5$ below the turnover, where the flux density $S_{\nu}$ at a frequency $\nu$ is proportional to ${\nu}^{\alpha}$. This replicates the empirical inverse power-law dependence of rest-frame turnover frequency on linear size~\citep{1997AJ....113..148O}, and results in magnetic field strengths consistent with equipartition estimates for GPS and HFP sources~\citep{2008A&A...487..885O}. In comparison, free-free absorption (FFA) of photons in a sufficiently dense external medium could also cause spectral turnover. Whereas FFA by an internal ionized screen has a characteristic optically thick index of $\alpha_{thick}{\sim}2.1$~\citep{2015ApJ...809..168C}, external FFA can result in much steeper spectra depending on properties of the absorbing cloud~\citep{1997ApJ...485..112B}. Such variations of FFA can model a similar inverse power-law relationship between linear size and turnover frequency~\citep{1997ApJ...485..112B}, and FFA has been shown to be necessary to accurately model spectra of various peaked-spectrum  sources~\citep{2000PASJ...52..209K,2001ApJ...550..160M,2005ASPC..347...29T,2015AJ....149...74T,2015ApJ...809..168C}.

In addition to the various spectral classes, previous VLBI imaging of peaked-spectrum sources has identified two distinct morphological classes~\citep{1997A&A...325..943S}. Core-jet structures are typically one sided with a steep spectrum jet and flat spectrum core, lacking the double morphology of larger radio doubles. On the other hand, symmetric structures are two sided with radio emission dominated by lobes and can include a weaker core component. Such symmetric objects have attracted interest appearing to be progenitors of FR I and II AGN~\citep{2014ApJ...780..178M}. Symmetric objects are further classified based on their size as compact symmetric objects (CSO, $\lesssim$1 kpc) or medium-sized symmetric objects (MSO, $\gtrsim$1 kpc)~\citep{2001A&A...369..380F}. 

With the introduction of advanced low frequency telescopes such as the LOw-Frequency ARray (LOFAR,~\citealt{2013A&A...556A...2V}), the Giant Metrewave Radio Telescope (GMRT,~\citealt{1991ASPC...19..376S}), and the Murchison Widefield Array (MWA,~\citealt{2013PASA...30....7T}), astronomers can now readily identify peaked-spectrum sources which have observed spectral peaks at megahertz frequencies. It is unclear how these MPS sources fit into the wider peaked-spectrum source population. Within the GaLactic and Extragalactic All-sky Murchison Widefield Array (GLEAM,~\citealt{2015PASA...32...25W, 2017MNRAS.464.1146H}) survey,~\cite{2017ApJ...836..174C} identified 1483 sources with spectral turnovers at frequencies from 72 MHz to 1.4 GHz. Some of these MPS sources were found to have high optically thick spectral indices in clear violation of SSA. The implication of this steep spectrum below the turnover on linear size and radio morphology is not yet understood.

VLBI is an invaluable tool in order to determine how properties of MPS sources compare to the wider peaked-spectrum source population. European VLBI Network (EVN) observations of 11 MPS sources suggested they were likely CSS, GPS, and HFP sources whose peak had been redshifted to megahertz frequencies through cosmological expansion~\citep{2016MNRAS.459.2455C}, however it is important to conduct multi-frequency VLBI imaging of \textit{bona-fide} MPS sources with spectra well sampled below the observed turnover. Multi-frequency VLBI analysis is critical for morphological classification in order to distinguish core-jet structures with flat spectrum cores from symmetric objects with two-sided jets. Moreover, multi-frequency VLBI can help determine spectral indices of sub-components in order to make equipartition magnetic field estimates, and increased resolution at higher frequencies can more accurately constrain linear size estimates. The purpose of this paper will be to use Very Long Baseline Array (VLBA) observations of 6 MPS sources identified by \cite{2017ApJ...836..174C} to constrain their morphology, linear size, and magnetic fields with the aim of understanding the nature of MPS sources and how they fit within the AGN evolution paradigm.

In Section~\ref{Sec:Observations} of this paper we detail our selection criteria, outline our observations, and describe our data calibration procedure. In Section~\ref{Sec:Results}, we present a description of targeted sources and results from synthesis imaging. We discuss source properties derived from our results and implications of our findings in Section~\ref{Sec:Discsussion}. Finally, we summarize our findings in Section~\ref{Sec:Conclusions}. A flat, Lambda Cold Dark Matter ($\Lambda$CDM) cosmological model with Hubble constant $H_{0}=$ 70 km s$^{-1}$ Mpc$^{-1}$ and density parameters $\Omega_{\textrm{M}}=$ 0.28 and $\Omega_{\Lambda}=$ 0.72~\citep{2013ApJS..208...19H} is adopted in this paper.


\begin{table*}
\caption{Source information and spectral characteristics~\citep{2017ApJ...836..174C} for the selected sample.}             
\label{Table:Sources}      
\centering 
\newcolumntype{C}{ @{}>{${}}c<{{}$}@{} }
\begin{tabular}{l
                l 
                l 
                *4{rCl}
                l 
                l}
\hline
\multicolumn{1}{c}{Source} & \multicolumn{1}{c}{R.A. (J2000,} & \multicolumn{1}{c}{Dec. (J2000,} & \multicolumn{3}{c}{${\nu}_{peak}$} & \multicolumn{3}{c}{S$_{peak}$} & \multicolumn{3}{c}{${\alpha}_{thick}$} & \multicolumn{3}{c}{${\alpha}_{thin}$} & \multicolumn{1}{c}{Observation,} & \multicolumn{1}{c}{$z$} \\
\multicolumn{1}{c}{      } & \multicolumn{1}{c}{hh mm ss.sss)} & \multicolumn{1}{c}{dd mm ss.ss)} & \multicolumn{3}{c}{(MHz)         } & \multicolumn{3}{c}{(Jy)      } & \multicolumn{3}{c}{                  } & \multicolumn{3}{c}{                 } & \multicolumn{1}{c}{Calibrator  } & \multicolumn{1}{c}{   } \\
\hline
J0231$-0433$ & 02 31 59.286 &  $-04$ 33 57.12 & 272 & \! \pm \! \! & 36 & 0.33 & \! \pm \! \! & 0.06  & 3.5 & \! \pm \! \! & 1.2 & $-0.57$  & \! \pm \! \! & 0.16   & A, J0228$-0337$ & 0.188   \\
J0235$-0100$ & 02 35 16.809 &  $-01$ 00 51.66 & 314 & \! \pm \! \! & 79 & 0.27  & \! \pm \! \! & 0.11   & 3.6 & \! \pm \! \! & 1.9 & $-0.34$  & \! \pm \! \! & 0.34   & A, J0239$-0234$ & 0.253   \\
J0330$-0740$ & 03 30 23.115 &  $-07$ 40 52.56 & 296 & \! \pm \! \! & 15 & 0.48 & \! \pm \! \! & 0.04  & 3.4 & \! \pm \! \! & 1.5 & $-0.23$ & \! \pm \! \! & 0.07 & A, J0335$-0709$ & 0.672   \\
J1509+1406 & 15 09 15.626 &  $+14$ 06 14.55 & 128 & \! \pm \! \! & 11 & 2.86 & \! \pm \! \! & 0.07  & 2.6 & \! \pm \! \! & 1.1 & $-0.90$ & \! \pm \! \! & 0.07 & B, J1507+1236 & ... \\
J1525+0308 & 15 25 48.957 &  $+03$ 08 25.95 & 112 & \! \pm \! \! & 30 & 5.51  & \! \pm \! \! & 0.21   & 3.2 & \! \pm \! \! & 2.0 & $-0.45$  & \! \pm \! \! & 0.12   & B, J1521+0420 & ... \\
J1548+1320 & 15 48 52.740 &  $+13$ 20 56.07 & 141 & \! \pm \! \! & 59 & 0.75 & \! \pm \! \! & 0.04  & 3.2 & \! \pm \! \! & 1.3 & $-0.79$ & \! \pm \! \! & 0.09 & B, J1553+1256 & ... \\
\hline                  
\end{tabular}
\tablefoot{Column 1: source name (J2000). 2: right ascension. 3: declination. 4: observed frequency of spectral peak. 5: flux density of spectral peak. 6: the spectral index ${\alpha}_{thick}$ in the optically thick regime derived from a generic curve fit~\citep{2017ApJ...836..174C}. 7: the spectral index ${\alpha}_{thin}$ in the optically thin regime derived from a generic curve fit. 8: observation (see Section~\ref{Sec:Observations}) and phase calibrator. 9: redshift (\citealt{1989AuJPh..42..633S, 2011ApJS..193...29A, 2009MNRAS.399..683J}; `...' if unknown).}
\end{table*}

\section{Sample Selection, Observations, and Data Reduction} \label{Sec:Observations}
\subsection{Sample Selection}
From the MPS sources identified by \cite{2017ApJ...836..174C} using the GLEAM survey, we considered a source a potential target if it had a declination $\delta > -8^{\circ}$ and a flux density above 0.1\,Jy at 1.4\,GHz in the National Radio Astronomy Observatory (NRAO) Very Large Array Sky Survey (NVSS,~\citealt{1998AJ....115.1693C}). Such constraints were to ensure reliable image reconstruction with the VLBA. We further required a subset of the target sources to appear to violate SSA based on the MWA-derived spectral indices in the optically-thick regime $\alpha_{thick}{\gtrsim}$2.5, and required such sources to have known redshifts in the literature. The second subset of targeted sources were required to have a clear spectral peak in the MWA band, between 72 and 230 MHz, but therefore had inadequate sampling of the optically thick regime to confirm SSA violation based on their spectra alone. We did not require this second subset to have known redshifts since our science goal was to explore, rather than comprehensively characterize, this new parameter space of milliarcsecond-scale structure of MPS sources. Properties of selected sources, J0231-0433, J0235-0100, J0330-0740, J1509+1406, J1525+0308, and J1548+1320, are reviewed in Table~\ref{Table:Sources}.


\subsection{Observations}
VLBA observations were carried out on 12 July 2017 and 13 July 2017 at the L and C-bands and the data are publicly available under project code SS008. Each band, centered at 1.55 GHz and 4.96 GHz, were divided into 8 sub-bands of 32 MHz width. We divided the observation into two parts: observation A which targeted the three sources with $\alpha_{thick}{\gtrsim}$2.5 and observation B which targeted the three sources with a spectral turnover between 72 and 230 MHz. Both observations included a 1 min gain calibrator observation (J0423-0120 for observation A, J1534+0131 for observation B) and observed targets for a total of ${\sim}$15 min in each band. To account for rapid phase variations, observations at 1.55 GHz cycled through 4 target scans of length 3:44 min and 5 phase calibrator scans of length 44 sec, and observations at 4.96 GHz cycled through 6 target scans of length 2:30 min and 7 phase calibrator scans of length 44 sec. Selected targets and respective calibrators used in phase referencing are listed in Table~\ref{Table:Sources}. Observation A included the Brewster, Kitt Peak, Los Alamos, Mauna Kea, North Liberty, Owens Valley, Pie Town, and St. Croix stations, while observation B additionally included the Hancock station. The Owens Valley station was selected as a reference antenna for both observations based on its superior post-flagging time and frequency coverage compared to other central antennae. 


\subsection{Data Reduction}
Post-correlation data reduction was conducted using the NRAO Astronomical Image Processing System (AIPS,~\citealt{AIPS}). First, edge channels and high amplitude spikes, mainly due to Global Positioning System signals, were flagged (UVFLG). Corrections for ionospheric dispersive delay were applied (VLBATECR), followed by corrections for errors in sampler thresholds (ACCOR), instrumental delay (PCCOR), and bandpass shape (VLBABPSS). System temperature and gain were used to calibrate amplitudes (APCAL) and phases were corrected for parallactic angle effects (VLBAPANG). For each target, phase solutions from fringe fitting (FRING) respective calibrators were applied. Remaining amplitude spikes were flagged (WIPER), and target field $u$, $v$, $w$ coordinates were recomputed and phase shifted (UVFIX) to correct for ${\sim}$arcsecond offsets from phase centers. 

Stokes I images with quasi-natural weighting (a~\citealt{Briggs} robustness of +1) were generated for all channels and sub-bands (IMAGR) using the Clark CLEAN algorithm \citep{1980A&A....89..377C} with a small number of components in tightly restricted regions to serve as an initial model for self calibration. Phase-only self calibration was then performed on the original data using entire scan lengths for solution intervals (CALIB). Targets were then re-imaged and cleaned with additional clean components in regions adjusted based on component placement, with components deemed unlikely based on placement plausibility or effect on final rms noise removed (CCEDT). Using this new model, phase-only self calibration on the original data based was repeated. This process of imaging with deeper cleaning in adjusted regions and phase-only self calibrating continued until the rms noise increased, which took between 2 and 7 iterations. Solution intervals were also decreased if this generated sensible solutions and lead to a reduction in rms noise. Once the noise floor was reached, based on the final model one round of phase and amplitude self calibration was performed on the last self-calibrated data using the entire scan length as a solution interval, if this reduced the rms noise.


\section{Results} \label{Sec:Results}

\begin{table}
\caption{Beam size and rms noise for images in Figures~\ref{Fig:ObservationA} and~\ref{Fig:ObservationB}.}             
\label{Table:Images}      
\centering          
\begin{tabular}{l l l l l}
\hline
\multicolumn{1}{c}{Source,} & \multicolumn{1}{c}{$\theta_{beam, \: maj}$} & \multicolumn{1}{c}{$\theta_{beam, \: min}$} & \multicolumn{1}{c}{P.A.       } & \multicolumn{1}{c}{$\sigma$ (mJy/}  \\
\multicolumn{1}{c}{Band   } & \multicolumn{1}{c}{(mas)                  } & \multicolumn{1}{c}{(mas)                  } & \multicolumn{1}{c}{ ($^\circ$)} & \multicolumn{1}{c}{beam)         }  \\
\hline
J0231-0433, L & 18.8 & 6.27 & $ 19.6$ & 0.727 \\
J0231-0433, C & 5.60 & 1.80 & $ 21.0$ & 0.296 \\
J0235-0100, L & 17.3 & 8.53 & $ 6.41$ & 0.213 \\
J0235-0100, C & 5.62 & 1.75 & $ 20.5$ & 0.394 \\
J0330-0740, L & 20.7 & 6.27 & $ 15.7$ &  2.13 \\
J0330-0740, C & 6.28 & 1.57 & $ 19.9$ & 0.491 \\
J1509+1406, L & 12.2 & 4.64 & $ 0.78$ & 0.827 \\
J1509+1406, C & 3.89 & 1.19 & $-5.01$ & 0.096 \\
J1525+0308, L & 13.8 & 4.92 & $-6.52$ &  5.32 \\
J1525+0308, C & 3.67 & 1.14 & $13.34$ &  1.59 \\
J1548+1320, L & 10.2 & 3.30 & $-4.73$ & 0.360 \\
J1548+1320, C & 4.14 & 1.47 & $-4.98$ & 0.051 \\
\hline    
\end{tabular}
\tablefoot{Column 1: source name (J2000) and frequency band (C$\equiv$4.96 GHz, L$\equiv$1.55 GHz). 2: major axis of the the synthesized beam. 3: minor axis. 4: position angle. 5: image background rms noise.}
\end{table}

\begin{figure*} \includegraphics[width=0.3384017341040463\linewidth]{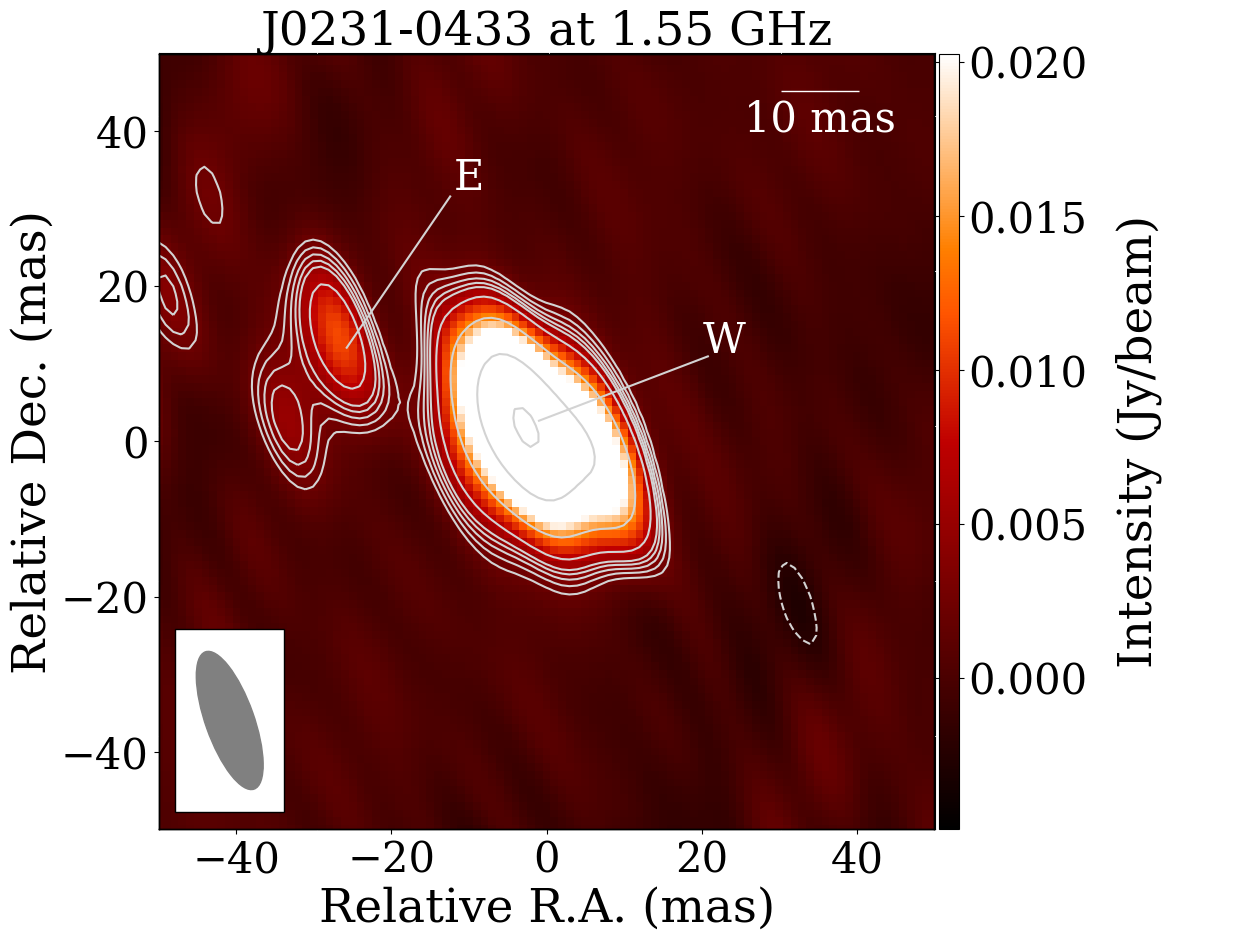}\quad\includegraphics[width=0.3384017341040463\linewidth]{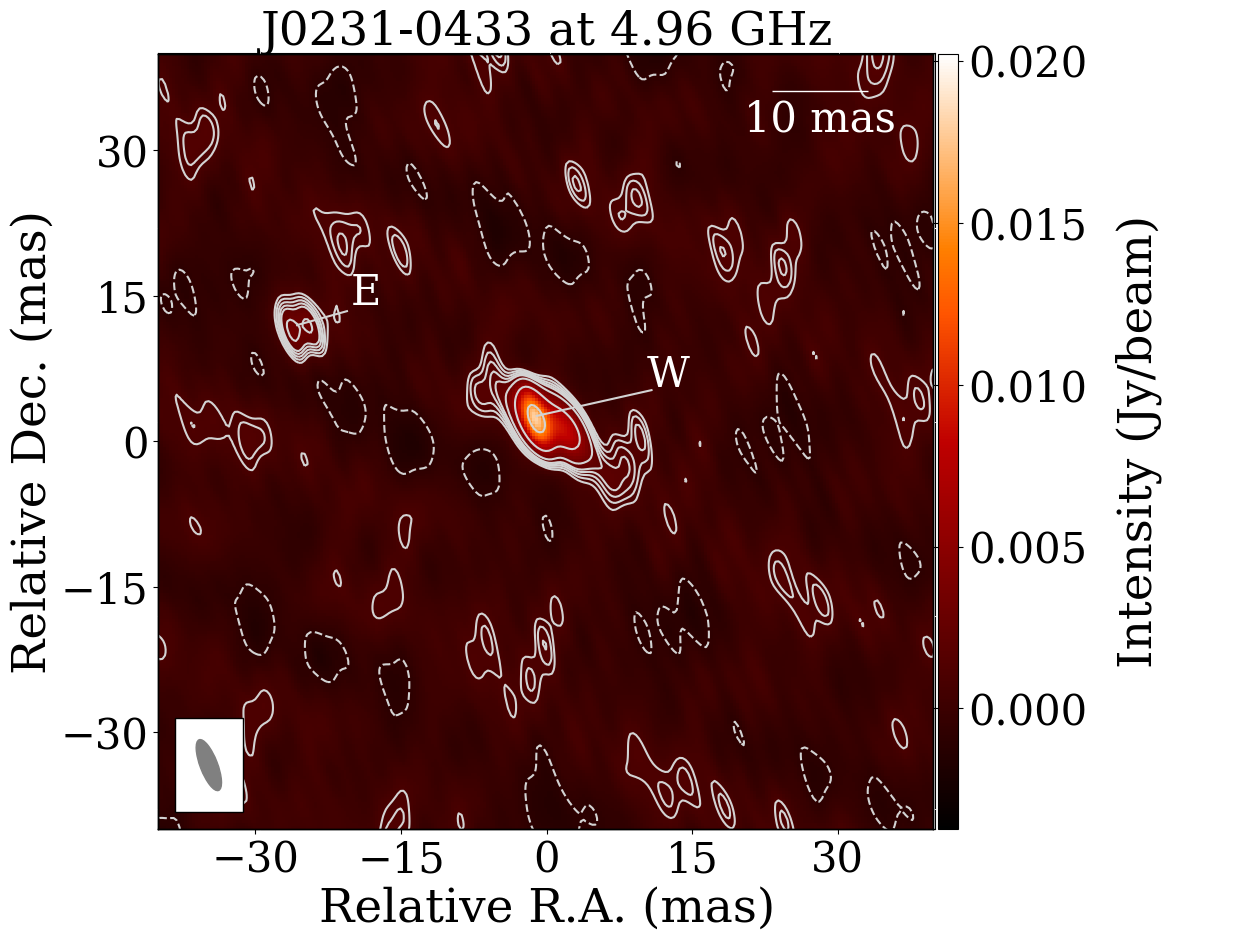}\quad\includegraphics[width=0.2531965317919075\linewidth]{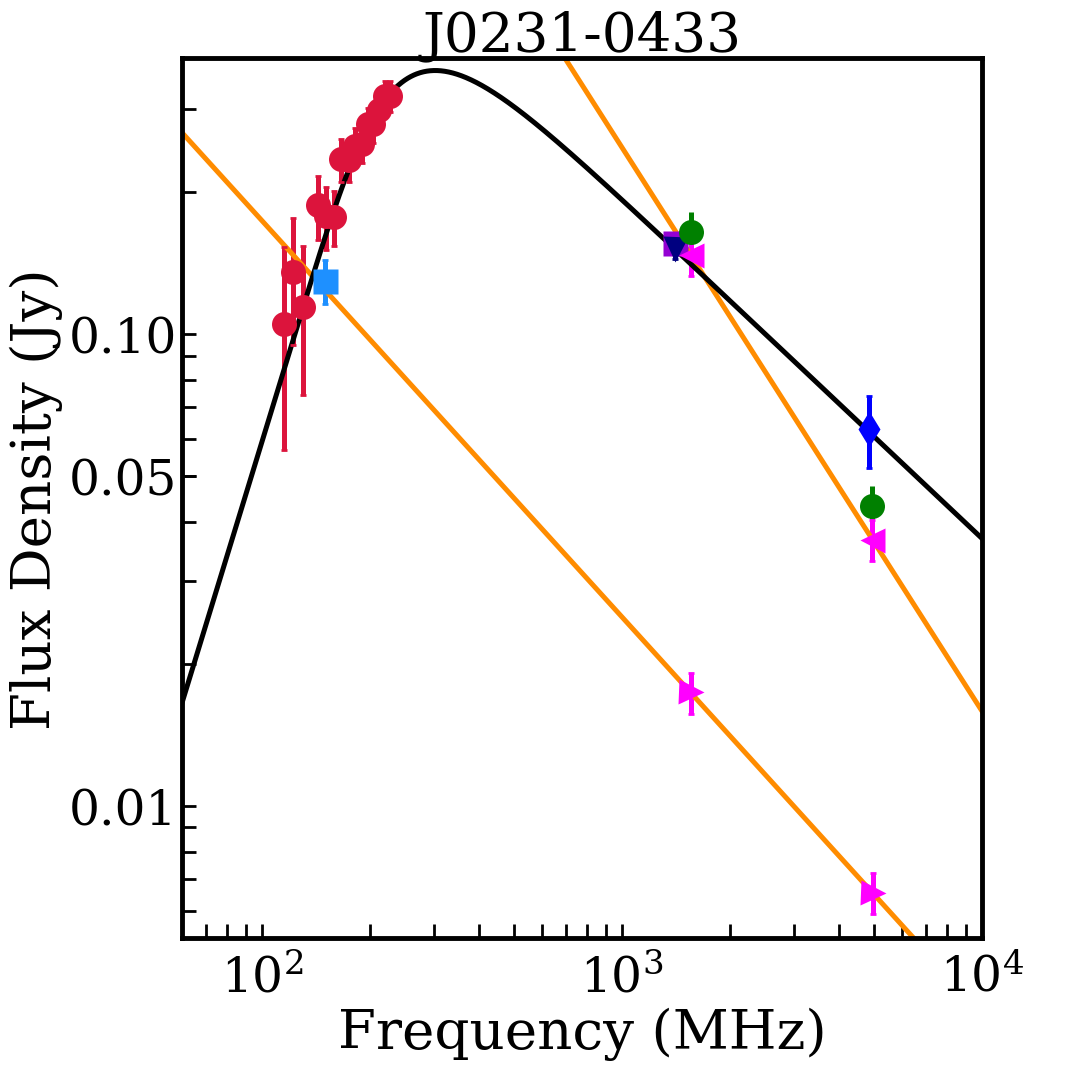} \\[\baselineskip]
                \includegraphics[width=0.3384017341040463\linewidth]{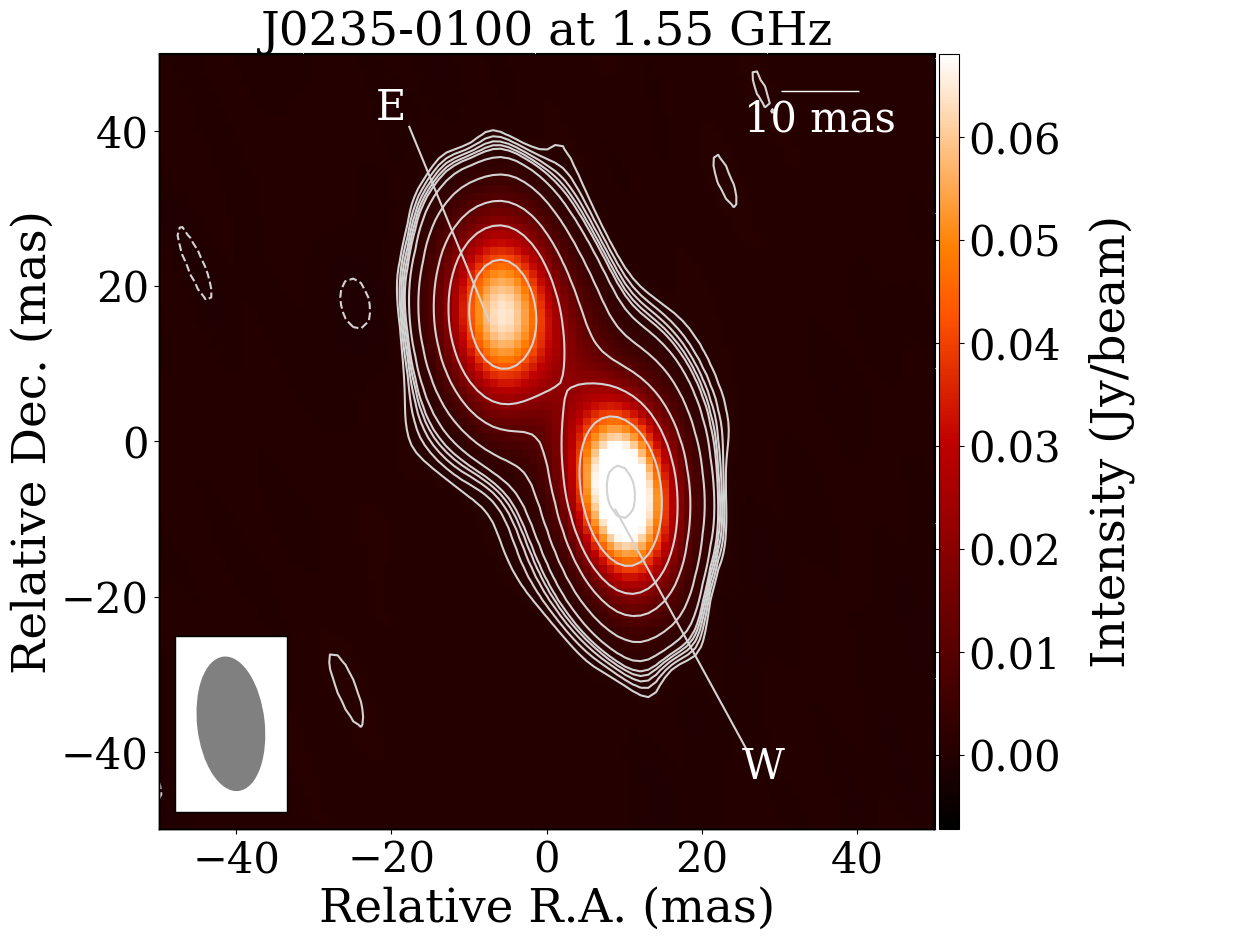}\quad\includegraphics[width=0.3384017341040463\linewidth]{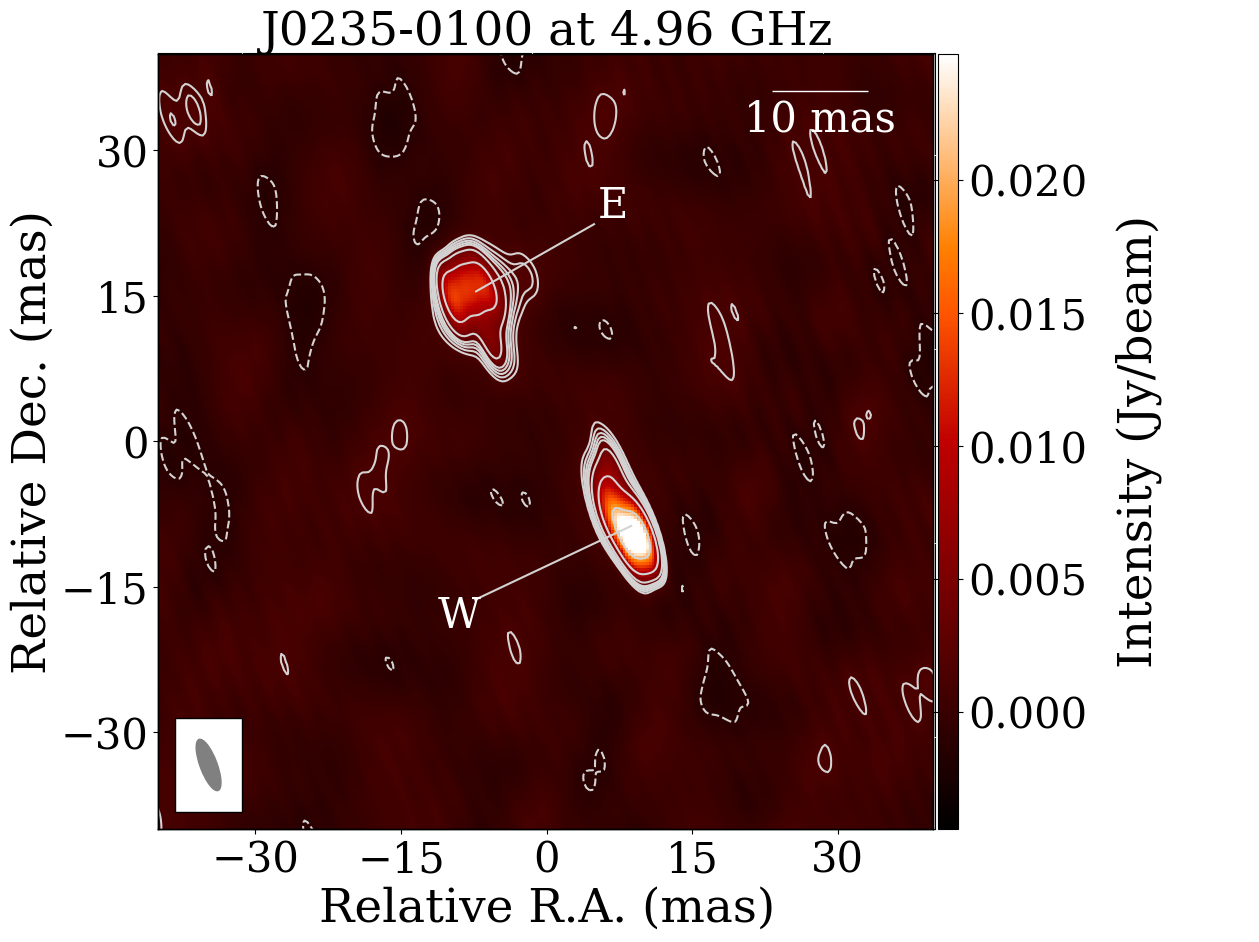}\quad\includegraphics[width=0.2531965317919075\linewidth]{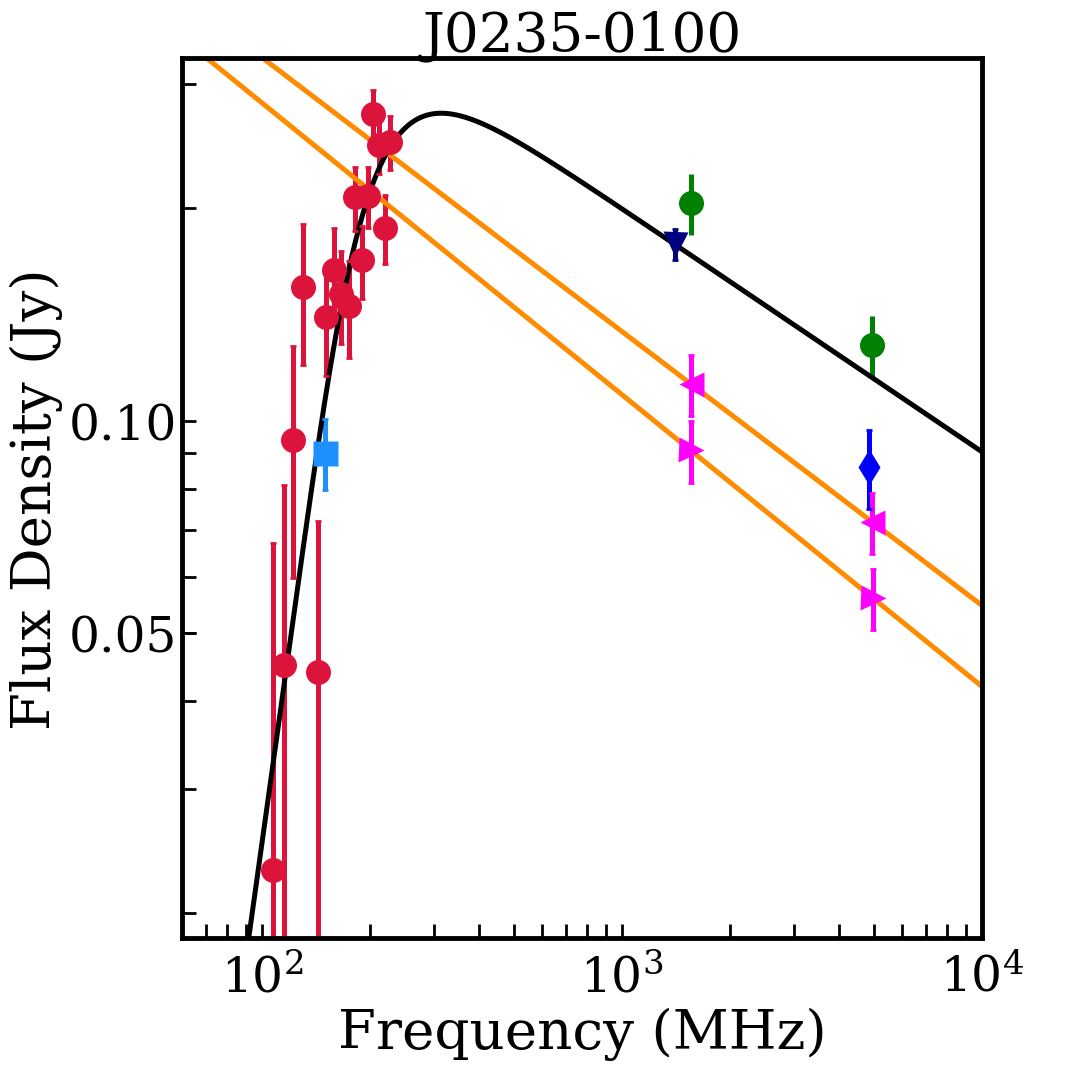} \\[\baselineskip]
                \includegraphics[width=0.3384017341040463\linewidth]{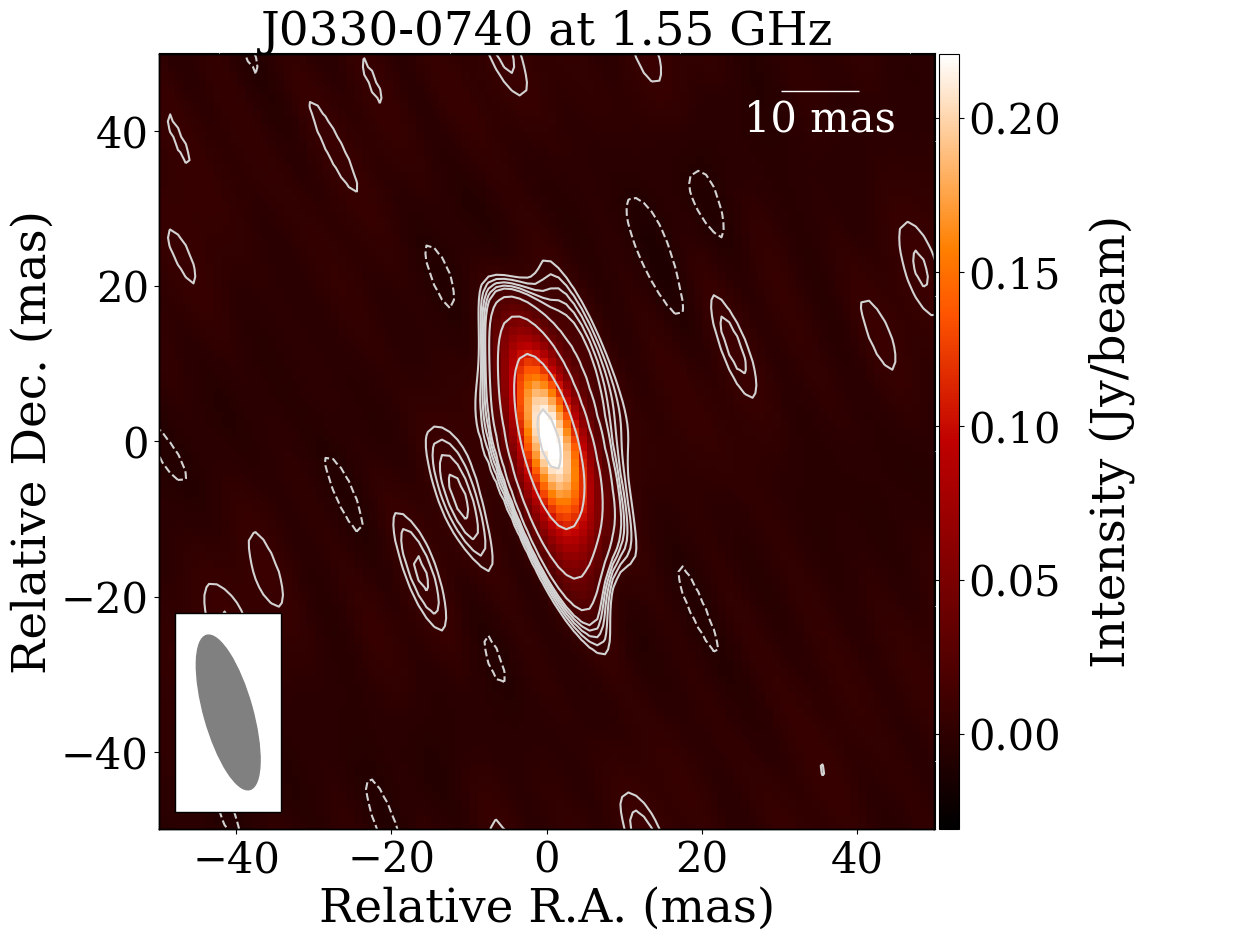}\quad\includegraphics[width=0.3384017341040463\linewidth]{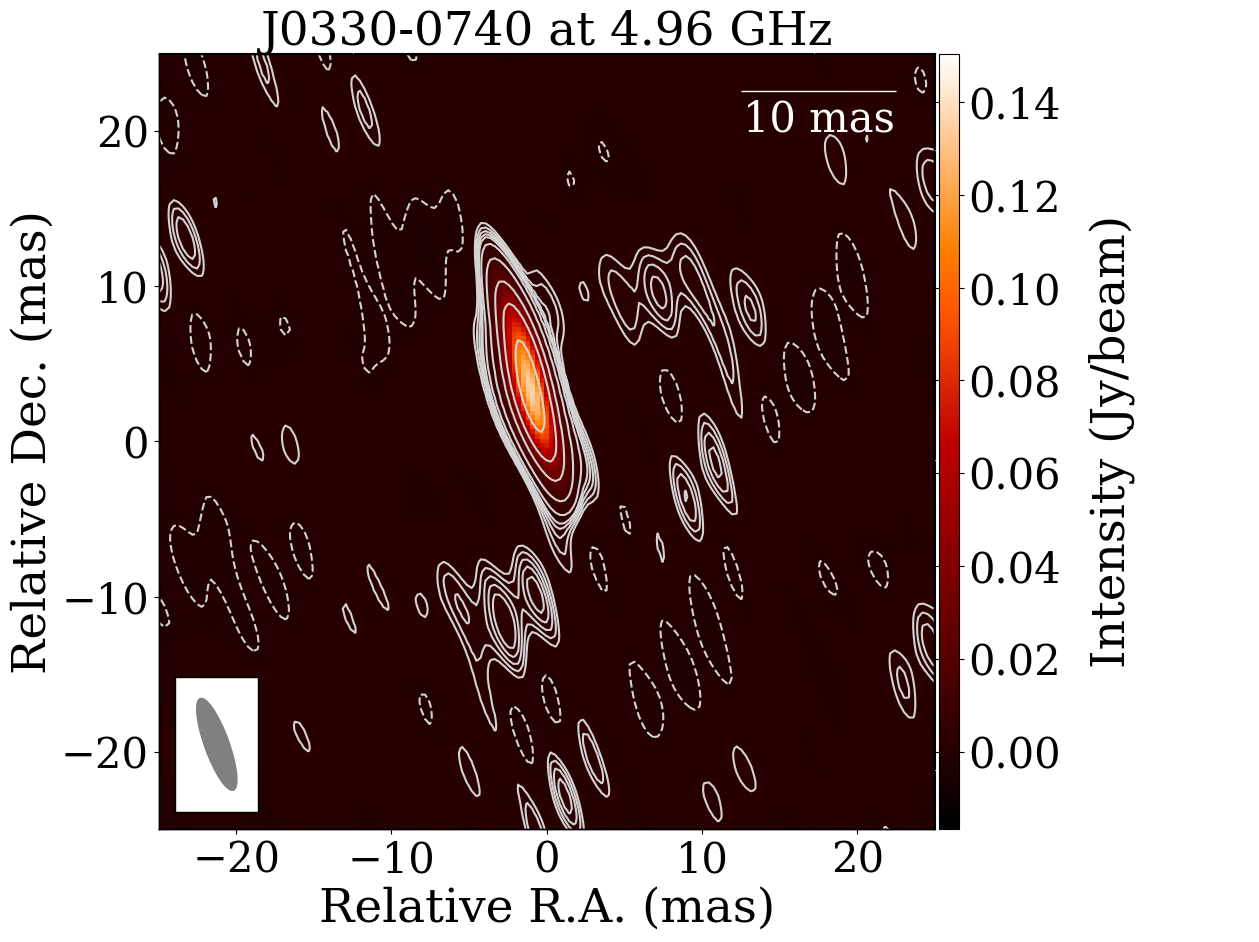}\quad\includegraphics[width=0.2531965317919075\linewidth]{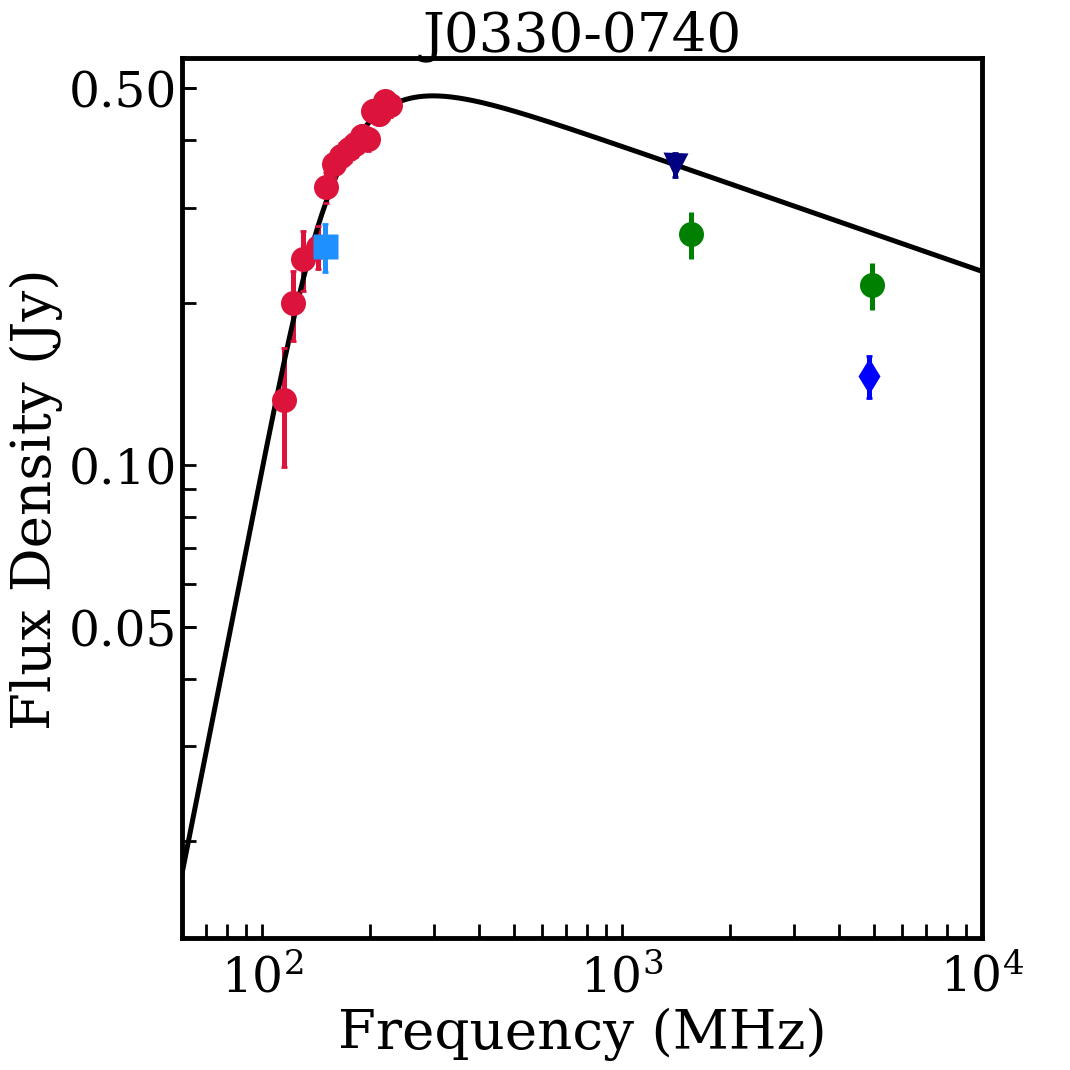} \caption{VLBA images at 1.55 GHz (left) and 4.96 GHz (middle) of sources with $\alpha_{thick}{\gtrsim}$2.5 and associated spectra (right). The Stokes I images have beam sizes and position angles as specified in Table~\ref{Table:Images}, with axes given in relative offset from R.A. and Dec. coordinates reported in Table~\ref{Table:Sources}. Contours are placed at ($-$3, 3, 4, 5, 6, 7, 10, 20, 50, 100, 200, 400, 800, 1600) $\times$ $\sigma$ (with $\sigma$ given in Table~\ref{Table:Images}). Color is given in a linear scale as indicated by color-bars to the right of the images. Where source components are resolved, Eastern and Western components are labeled as `E' and `W' at the same R.A. and Dec. for both images based on component peak locations at 4.96 GHz. Spectra include data from GLEAM in red circles, TGSS-ADR1 in blue squares, NVSS in navy downward-pointing triangles, FIRST in dark violet squares, and PMN in dark blue diamonds. Integrated flux densities from associated VLBA images are included as green circles, and, where sources are resolved, Eastern and Western components are included as right and left pointing magenta triangles, respectively. The black curve indicates the fit of a generic curve to GLEAM, TGSS-ADR1, and NVSS data. Where sources are resolved at both frequencies, orange lines represent the resulting power-law for each component.} \label{Fig:ObservationA} \end{figure*}
\begin{figure*} \includegraphics[width=0.3384017341040463\linewidth]{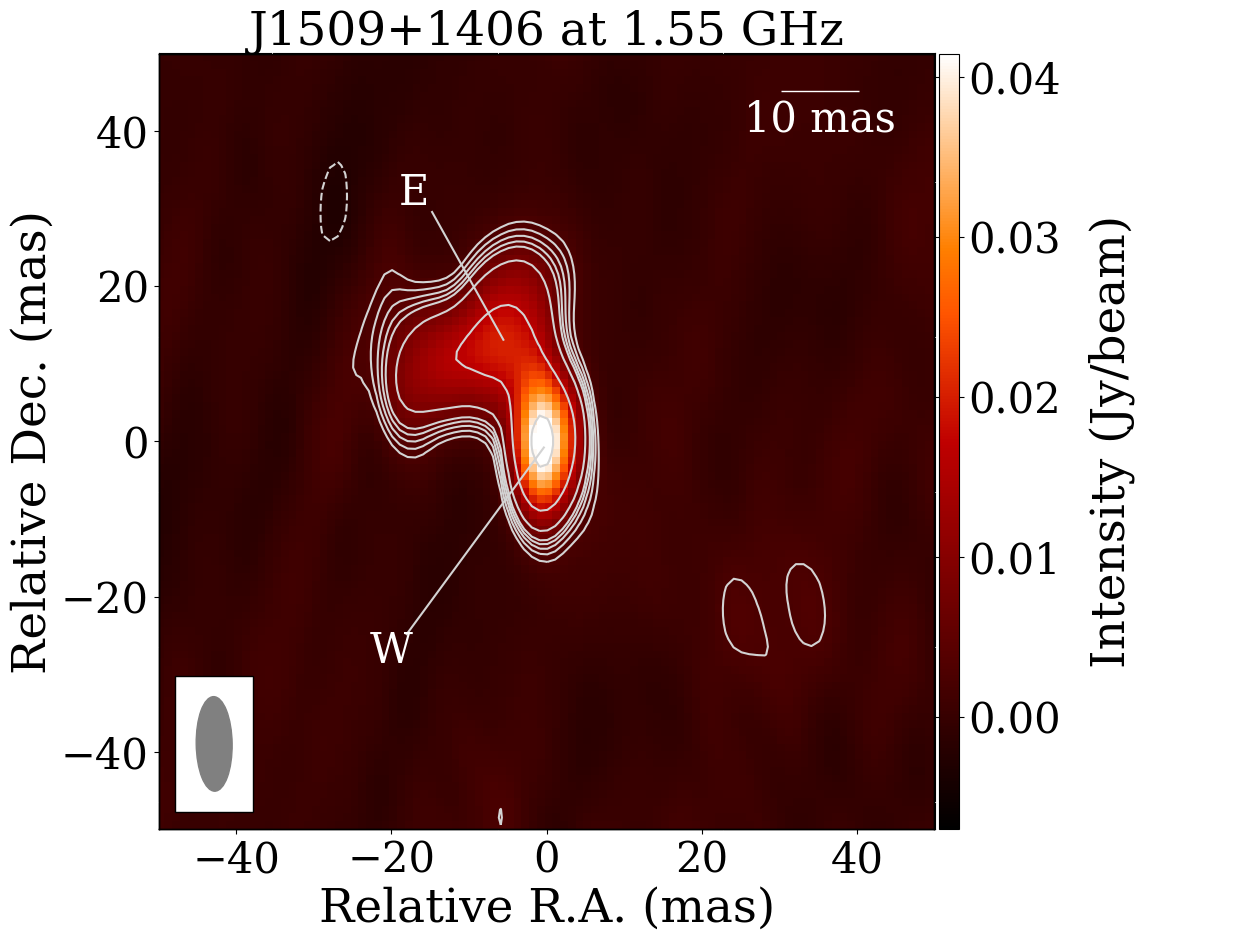}\quad\includegraphics[width=0.3384017341040463\linewidth]{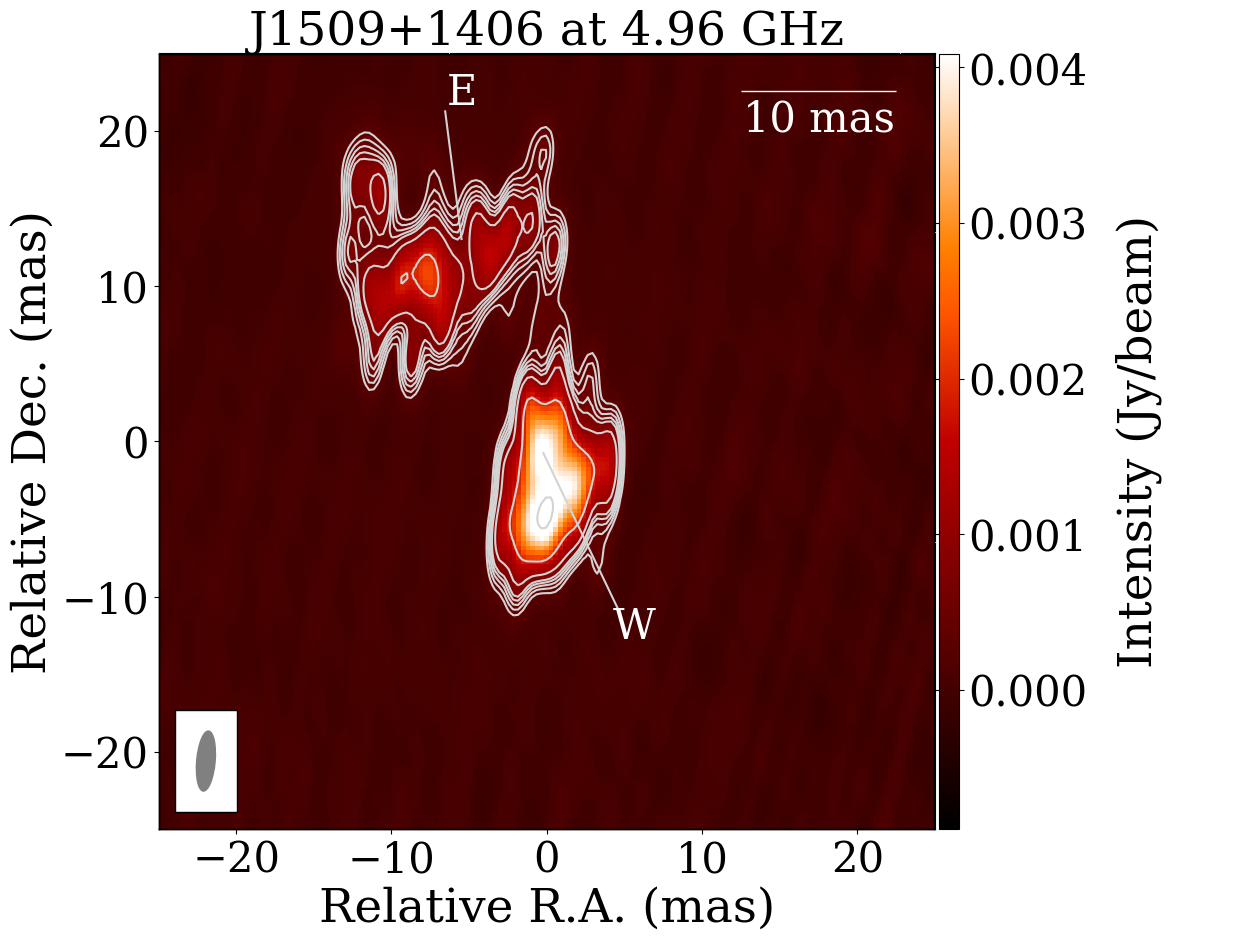}\quad\includegraphics[width=0.2531965317919075\linewidth]{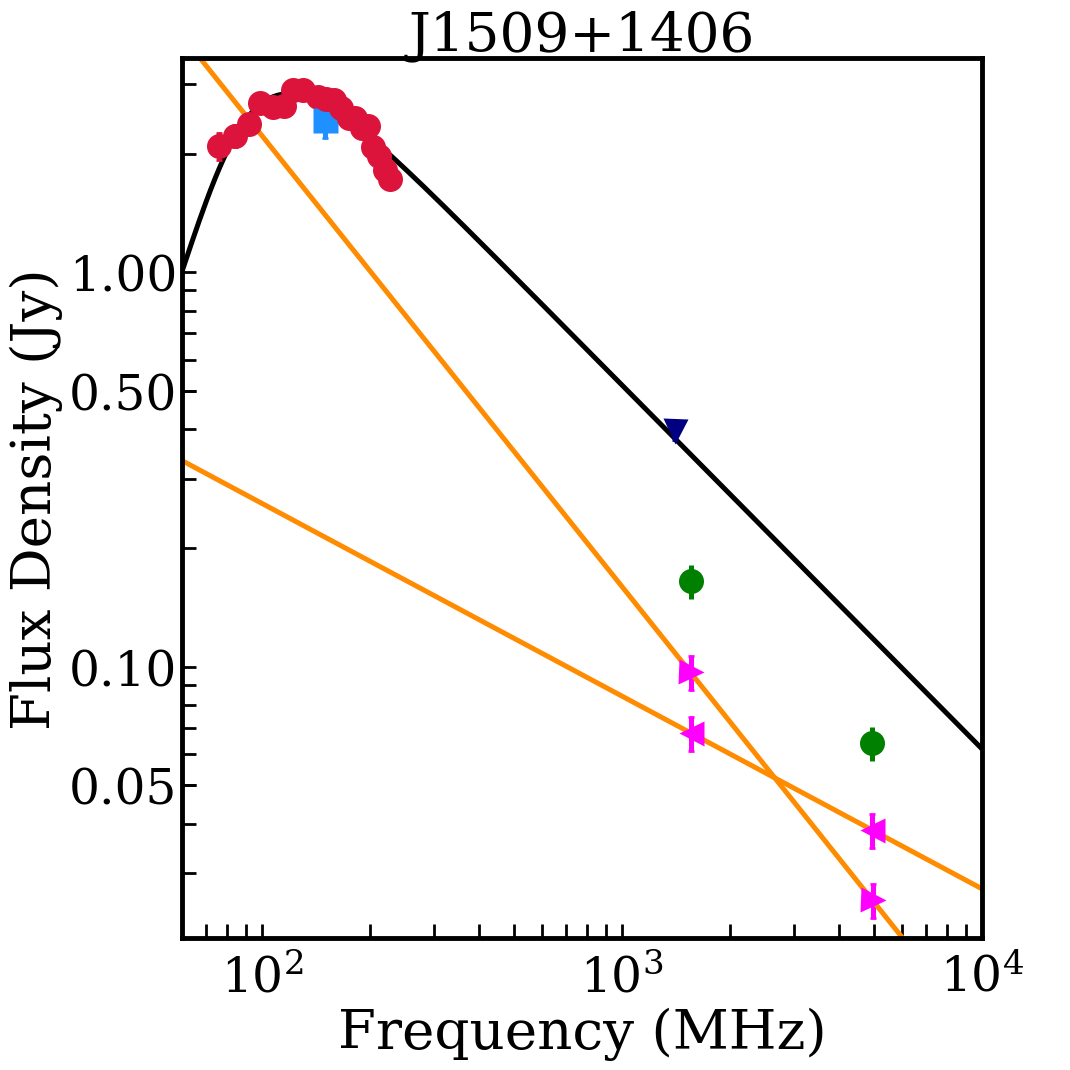} \\[\baselineskip]
                \includegraphics[width=0.3384017341040463\linewidth]{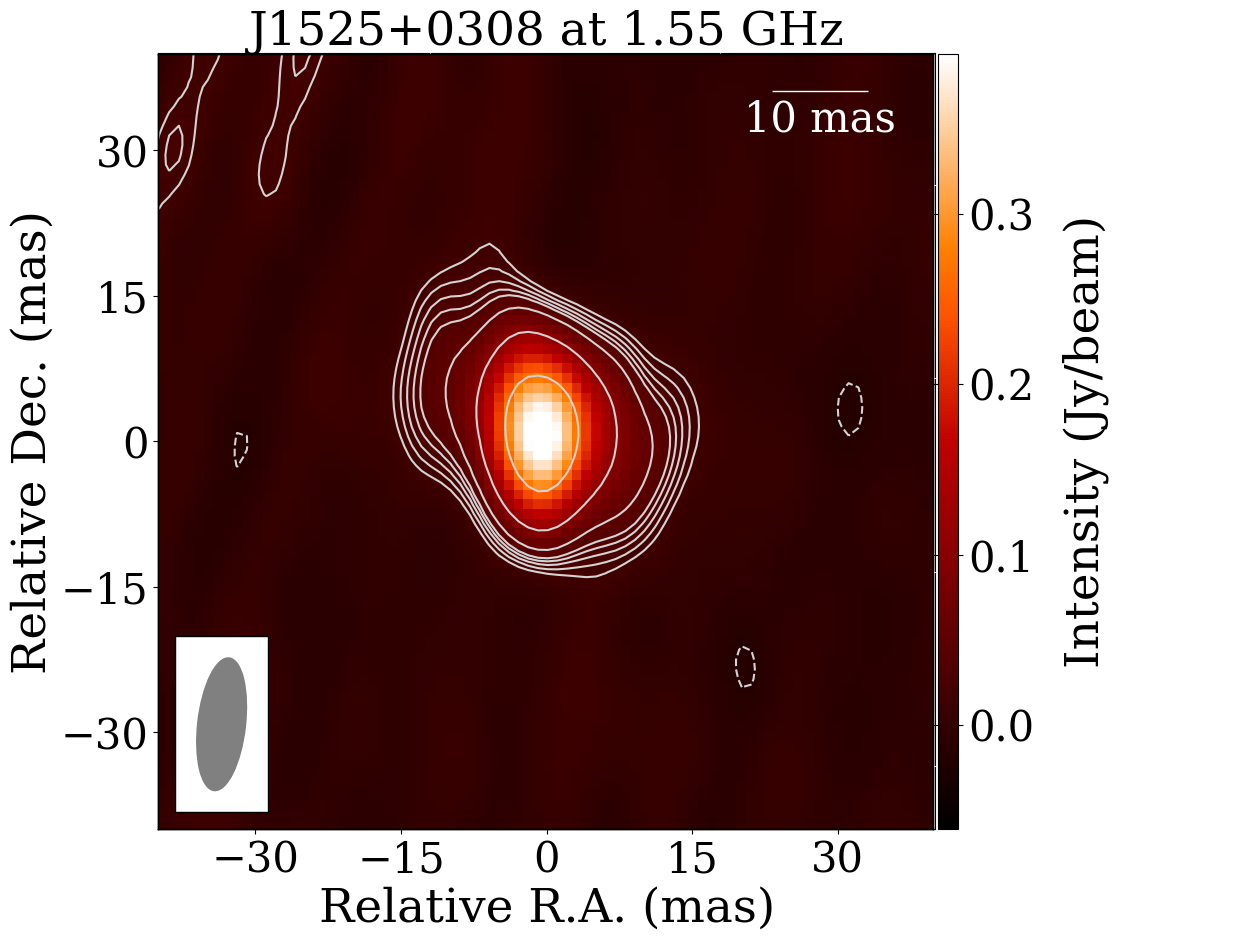}\quad\includegraphics[width=0.3384017341040463\linewidth]{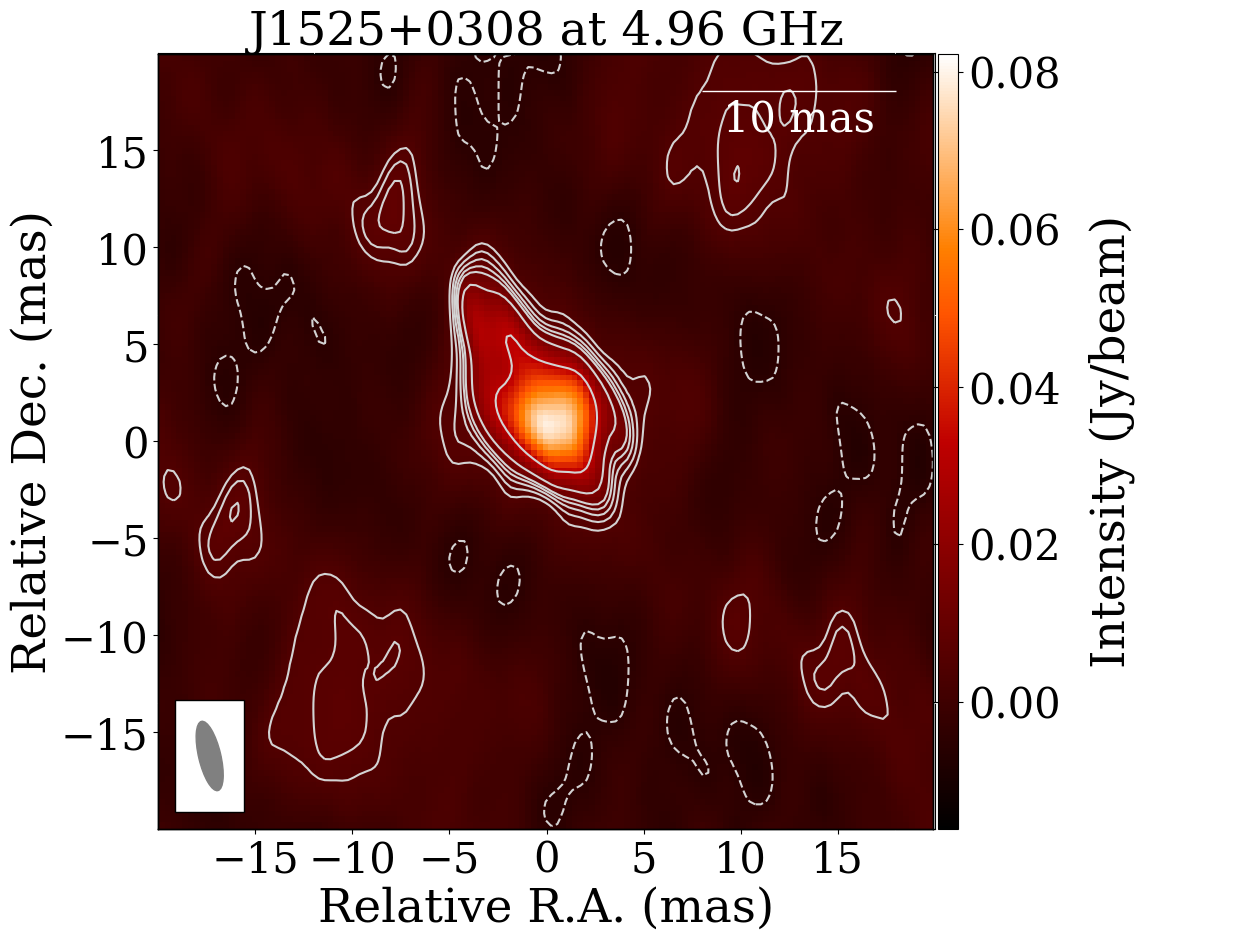}\quad\includegraphics[width=0.2531965317919075\linewidth]{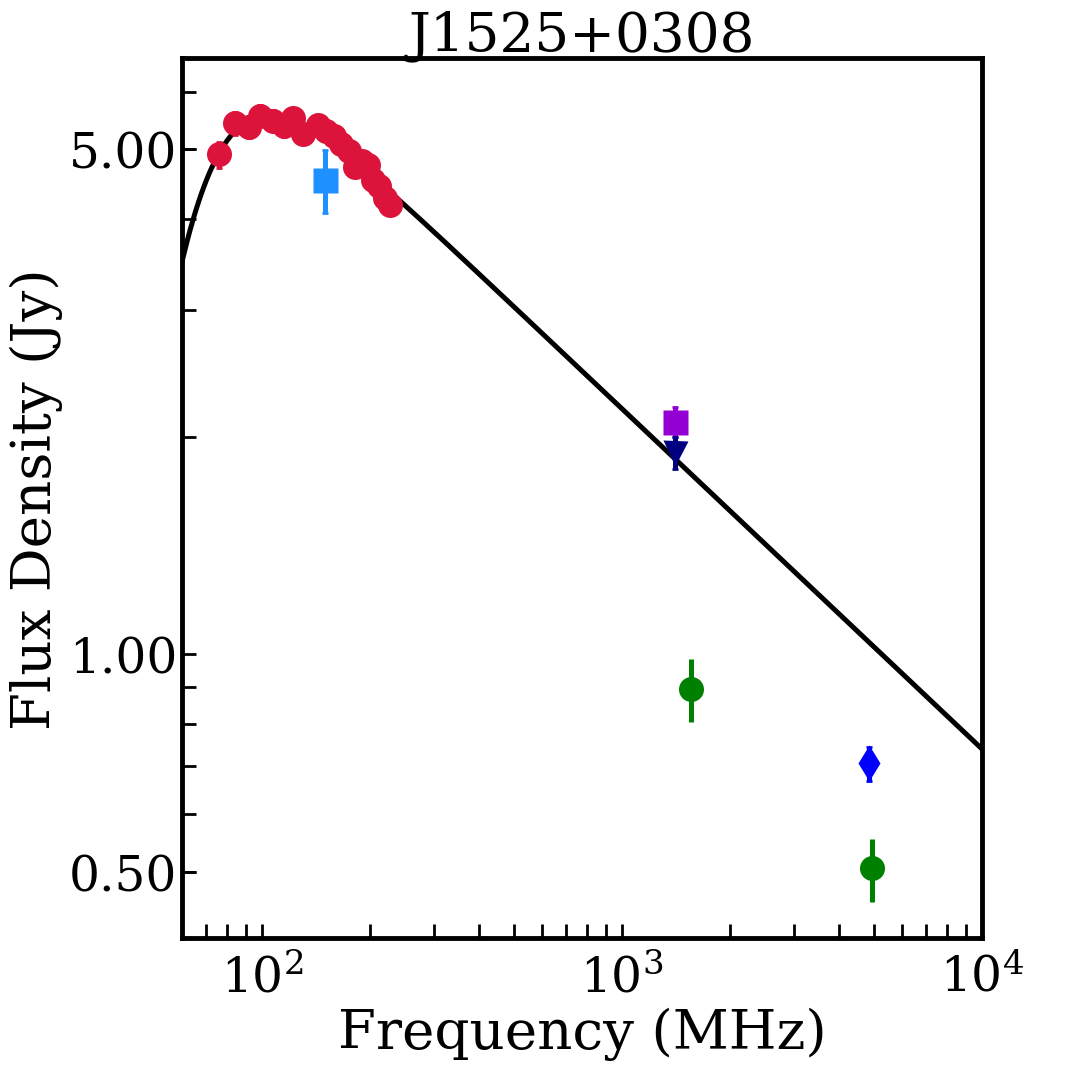} \\[\baselineskip]

                \includegraphics[width=0.3384017341040463\linewidth]{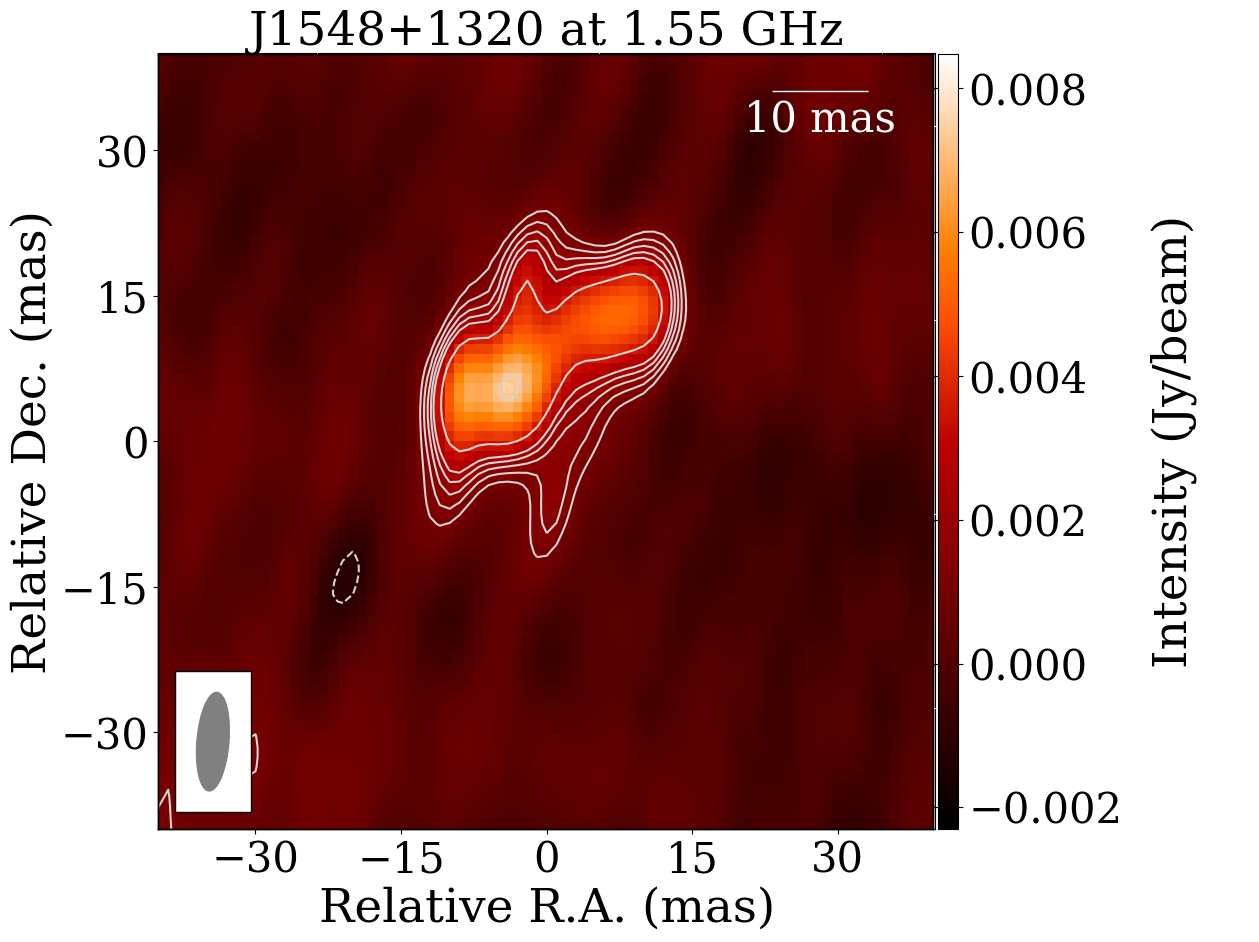}\quad\includegraphics[width=0.3384017341040463\linewidth]{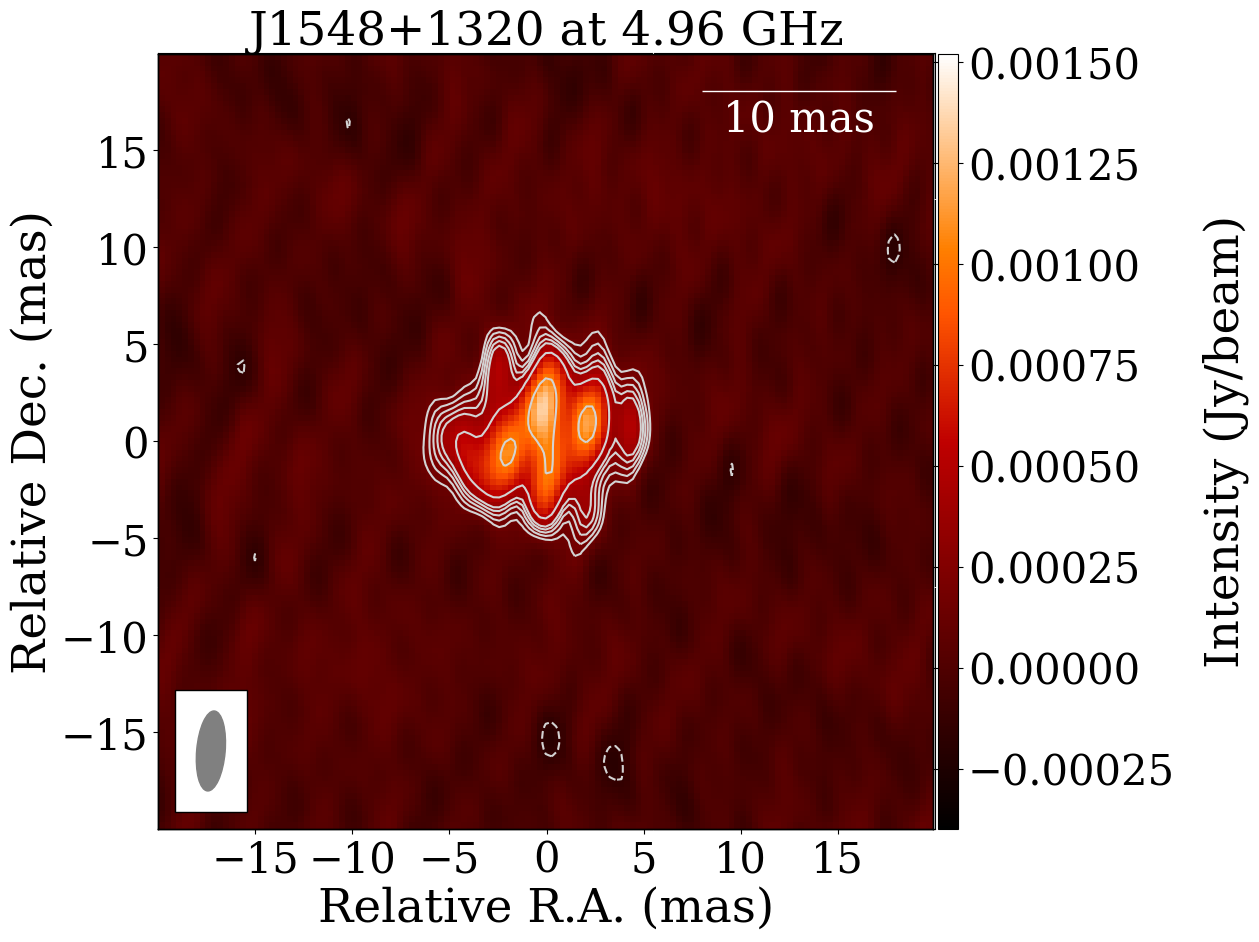}\quad\includegraphics[width=0.2531965317919075\linewidth]{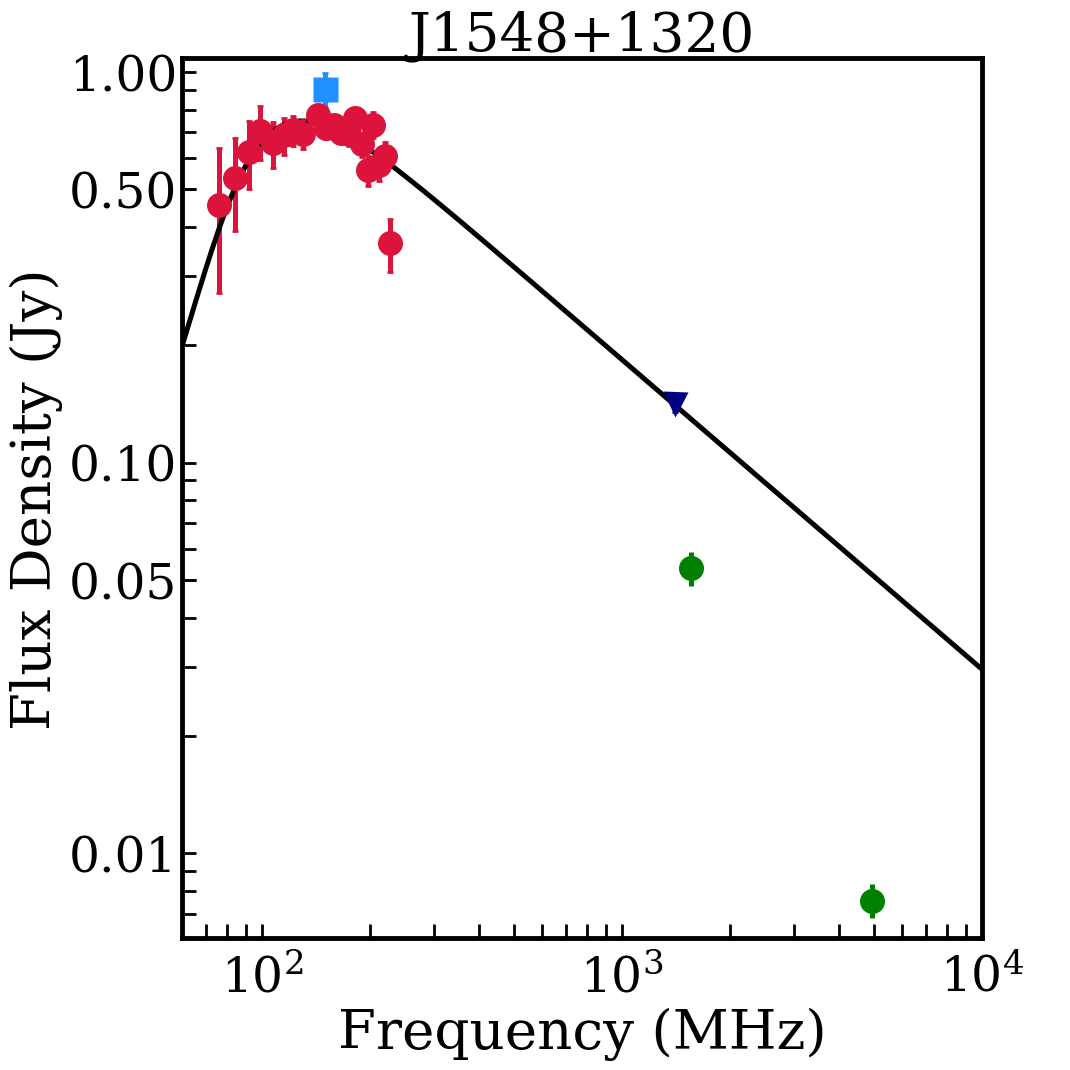} \caption{VLBA images at 1.55 GHz (left) and 4.96 GHz (middle) of sources with peaks between 72 and 230 MHz and associated spectra (right). The Stokes I images have beam sizes and position angles as specified in Table~\ref{Table:Images}, with axes given in relative offset from R.A. and Dec. coordinates reported in Table~\ref{Table:Sources}. Contours are placed at ($-$3, 3, 4, 5, 6, 7, 10, 20, 50, 100, 200, 400, 800, 1600) $\times$ $\sigma$ (with $\sigma$ given in Table~\ref{Table:Images}). Color is given in a linear scale as indicated by color-bars to the right of the images. Where source components are resolved, Eastern and Western components are labeled as `E' and `W' at the same R.A. and Dec. for both images based on component peak locations at 4.96 GHz. Spectra include data from GLEAM in red circles, TGSS-ADR1 in blue squares, NVSS in navy downward-pointing triangles, FIRST in dark violet squares, and PMN in dark blue diamonds. Integrated flux densities from associated VLBA images are included as green circles, and, where sources are resolved, Eastern and Western components are included as right and left pointing magenta triangles, respectively. The black curve indicates the fit of a generic curve to GLEAM, TGSS-ADR1, and NVSS data. Where sources are resolved at both frequencies, orange lines represent the resulting power-law for each component.} \label{Fig:ObservationB} \end{figure*}

We present 1.55 GHz and 4.96 GHz images in Figures~\ref{Fig:ObservationA} and~\ref{Fig:ObservationB} for sources with $\alpha_{thick}{\gtrsim}$2.5 and sources with peaks between 72 and 230 MHz, respectively. Synthesized beam size and rms noise for the images are outlined in Table~\ref{Table:Images}. In the interest of simple reproducibility we calculated source and component integrated flux densities for both 1.55 GHz and 4.96 GHz images using BANE~\citep{2018PASA...35...11H} and Aegean~\citep{2012MNRAS.422.1812H} to estimate background and noise properties, identify component pixel groups, sum over pixels, and divide by the synthesized beam, replicating traditional integration over 3$\sigma$ isophotes. Where sources are resolved at both frequencies, for each component we find the resulting $\alpha$ to satisfy the power law $S_{\nu}\propto{\nu}^{\alpha}$. Following~\cite{2017ApJ...836..174C}, we include a generic curved model fit \citep{1998A&AS..131..435S} to data from the GLEAM survey, the Tata Institute of Fundamental Research GMRT Sky Survey Alternative Data Release (TGSS-ADR1,~\citealt{2017A&A...598A..78I}), and the NVSS. Figures~\ref{Fig:ObservationA} and~\ref{Fig:ObservationB} also include data from the Faint Images of the Radio Sky at Twenty centimeters (FIRST,~\citealt{2015ApJ...801...26H}) survey and the Parkes-MIT-NRAO (PMN,~\citealt{1994ApJS...91..111W}) survey using general-width fits where possible~\citep{1993AJ....105.1666G}. We have estimated the relative error of flux density measurements (see Figures~\ref{Fig:ObservationA} and~\ref{Fig:ObservationB}) based on the distribution of average amplitude measurements per baseline for the gain calibrator, which was ${\sim}$10\%.

Here, we review morphological properties and spectral characteristics for each source. All dimensions, including component separation and linear sizes, are measured from the 4.96 GHz images.


\subsection{J0231-0433}
J0231-0433 (GLEAM J023159-043352) was resolved into a bright 5.9 $\times$ 3.8 milliarcsecond (mas) Western component and a faint 5.3 $\times$ 1.9 mas Eastern component separated by 26.5 mas. Component flux densities were found to be quite asymmetric, with the Western component ${\sim}$9 times brighter than the Eastern component at 1.55 GHz and ${\sim}$7 times brighter at 4.96 GHz. Total integrated flux density at 1.55 GHz was approximately equal to the generic curve fit, but 30$\pm$7\% of the flux density was missing compared to the curve fit at 4.96 GHz. It has a redshift $z$ of 0.188 assuming association with the galaxy cluster Abell 362~\citep{1989AuJPh..42..633S}, suggesting a linear size of 84 pc. It has an X-ray counterpart~\citep{2018yCat.9052....0C} and a Wide-field Infrared Survey Explorer (WISE,~\citealt{2010AJ....140.1868W}) counterpart J023159.27-043356.9, with colors suggesting it is either a luminous infrared galaxy or an obscured AGN. It has no optical counterpart and appears as an empty field in the Sloan Digital Sky Survey (SDSS,~\citealt{2019ApJS..240...23A}). The steep spectral indices of both components (${\alpha}_{W}{\sim}-1.2$, ${\alpha}_{E}{\sim}-0.8$) suggest they are or include jets, making the source a CSO candidate. 


\subsection{J0235-0100}
J0235-0100 (GLEAM J023516-010051) was resolved into a 7.0 $\times$ 3.0 mas Western component and a slightly more faint 5.7 $\times$ 4.4 mas Eastern component separated by 25.2 mas. Component flux densities were found to be comparable, with the Western component ${\sim}$1.2 times brighter than the Eastern component at 1.55 GHz and ${\sim}$1.3 times brighter at 4.96 GHz. It has as an optical counterpart SDSS J023516.80-010051.7, a galaxy at $z$=0.253~\citep{2011ApJS..193...29A} resulting in a linear size of 100 pc. It has an infrared counterpart from the Two Micron All Sky Survey (2MASS,~\citealt{2006AJ....131.1163S}) J02351685-0100512 and WISE J023516.81-010051.5, with colors suggesting it is most likely a luminous infrared galaxy. It has been classified both as a low and high excitation radio galaxy~\citep{2012MNRAS.421.1569B, 2017MNRAS.464.1306C}, however its radio power suggests the latter is more likely. While the components of J0235-0100 are not necessarily steep (${\alpha}_{W}{\sim}-0.39$, ${\alpha}_{E}{\sim}-0.41$), its symmetric structure suggests it is a CSO candidate. The 4.96 GHz image reveals even more complex structure in the Eastern component, which presents the possibility of a core identification given higher frequency VLBI study.


\subsection{J0330-0740}
We did not resolve J0330-0740 (GLEAM J033023-074052) and it was almost entirely confined to a 6.3 $\times$ 1.6 mas beam sized region at 4.96 GHz. 24$\pm$8\% of total integrated flux density was missing compared to the spectral model at 1.55 GHz, 20$\pm$8\% at 4.96 GHz. Its redshift was identified by the Six-degree Field Galaxy Survey (6dFGS,~\citealt{2009MNRAS.399..683J}) to be $z$=0.672, giving a projected linear size upper limit of 45 pc. It has an optical counterpart SDSS J033023.11-074052.6 and infrared counterparts 2MASS J03302312-0740524 and WISE J033023.12-074052.6, with colors suggesting it is a quasar.


\subsection{J1509+1406}
J1509+1406 (GLEAM J150915+140615, 4C +14.58) was resolved into a 14 $\times$ 6.2 mas Western component separated by 14.7 mas from a more complex Eastern component that might be a bent jet or a phenomenon caused by a unique viewing angle, broadly confined by a 18 $\times$ 16 mas region. Component flux densities were found to be quite comparable, with the Eastern component ${\sim}$1.4 times brighter than the Western component at 1.55 GHz, but the \textit{Western} component ${\sim}$1.5 times brighter than the Eastern component at 4.96 GHz. 52$\pm$5\% of total integrated flux density was missing compared to the curve fit at 1.55 GHz, 46$\pm$5\% at 4.96 GHz. The source has no optical or infrared counterparts and appears as an empty field in SDSS, WISE, and 2MASS. Lacking a known redshift, we take a $z$ of 1 to make an estimate of the linear size of $\ell {\sim}$120 pc.\footnote{The median redshift of radio galaxies with S$_{1.4 \: GHz}{\sim}$0.1 Jy is $z$=0.8 and the median for S$_{1.4 \: GHz}{\sim}$0.001 Jy galaxies is $z$=1~\citep{1984ApJ...287..461C}.}

The flat spectrum of the Western component (${\alpha}_{W}{\sim}-0.45$) compared to the Eastern component (${\alpha}_{E}{\sim}-1.1$) suggests the Western component is -- or at least contains -- the core, making it a core-jet candidate. While the spectrum of the `core' is not flat or inverted, components with optically thin spectral indices of $-0.7$ have been identified as cores (see J2136+0041,~\citealt{2006A&A...450..959O}). Additionally, it is possible that the Western component has substantial diffuse emission, meaning we have resolved out a considerable amount of flux density due to the lack of short baselines, and that its spectra is even flatter than displayed in Figure~\ref{Fig:ObservationB}. 

The identification of the Western component as a core is supported by the brightness temperature $T_{b}$ of the components
 \begin{equation}
      T_{b} \approx 1.22 \times 10^{12} \frac{S_{comp} \: (1+z)}{{\theta_{D, \: maj}} \: {\theta_{D, \: min}} \nu^2},\label{Equation:TB}
    \end{equation}
where $T_{b}$ is in K, $\theta_{D, \: maj}$ and $\theta_{D, \: min}$ are the effective deconvolved major and minor component axes in mas, and $S_{comp}$ is the flux density of a component in Jy at the observed frequency $\nu$ in GHz~\citep{2019A&A...622A..92N}. Following~\cite{2008A&A...487..885O} and~\cite{2016MNRAS.461.2879V}, we modify our size estimates by a factor of 1.8 to consider uniform sources and estimate the effective deconvolved size $\theta_{D}$ for major and minor axes of resolved components as
\begin{equation}
    \theta_{D} = 1.8 \times \sqrt{\theta_{comp}^2 - \theta_{beam, \: maj} \: \theta_{beam, \: min}} \: ,
\end{equation}
where $\theta_{comp}$ is either the major or minor axis of a component in mas. In a VLBI survey of 162 compact radio sources at 86 GHz,~\cite{2019A&A...622A..92N} found that core components had a median brightness temperature ${\sim}$12 times larger than that for jets. Since the brightness temperature of the Western component is ${\sim}$2 times that of the Eastern component at 1.55 GHz ($T_{b, W}=(2.7\pm0.6)\times10^{8}$ K compared to $T_{b, E}=(1.1\pm0.2)\times10^{8}$ K) and ${\sim}$5 times the Eastern component at 4.96 GHz ($T_{b, W}=(1.5\pm0.3)\times10^{7}$ K compared to $T_{b, E}=(2.8\pm0.6)\times10^{6}$ K), the comparative brightness temperatures would follow the trend at 86 GHz for `hotter' cores. 

However, since the interaction of a jet with a surrounding medium can cause high radiative losses and slow down the growth of jets leading to an underestimation of the source age~\citep{2016AN....337....9O}, it is possible that the comparatively sharp spectrum of the Eastern component may be a consequence of complex interaction with a medium, rather than suggesting the Western component to be the core. This is supported by the complex, seemingly bent nature of the jet.


\subsection{J1525+0308}
J1525+0308 (GLEAM J152548+030825, 4C +03.33) was not resolved into two components, but has interesting slightly resolved structure including extended emission towards the North-East as seen at 4.96 GHz. It is broadly confined by a 6.2 $\times$ 3.5 mas region. Due to diffuse emission, the source likely would have considerable power at baselines shorter than those included in the VLBA, and we find 50$\pm$5\% of total integrated flux density was missing compared to the curve fit at 1.55 GHz, 51$\pm$5\% at 4.96 GHz. It has an infrared counterpart WISE J152548.95+030826.1 with colors that suggest it is a quasar. It has faint counterpart in the SDSS field but no object entry. Based on the 5.4 mas separation of the center of the extended emission from the source's peak, we estimate an upper limit of the linear size to be 44 pc assuming $z$=1. Its flux density at 1.4 GHz was found not to vary by $\geq$4$\sigma$~\citep{2011ApJ...737...45O}.

We can further comment upon the likely redshift by comparison to the general radio source population explored in the HETDEX Spring Field by the LOFAR Two-metre Sky Survey First Data Release (LoTSS-DR1,~\citealt{2017A&A...598A.104S, 2019A&A...622A...1S}). While the source lacks a $K$-band magnitude to estimate a redshift from previous $K$-$z$ relations~\citep{2001qhte.conf..333J}, the relatively high sensitivity of the 3.4$\mu$m $W1$-band among the four $WISE$ bands makes it a good substitute to find a similar $W1$-$z$ relationship from the LoTSS-DR1 sources with known $W1$ magnitudes and redshifts~\citep{2019A&A...622A...2W, 2019A&A...622A...3D}. Given a $W1$-band magnitude of 14.78$\pm$0.03, we consider LoTSS-DR1 sources with $W1$-band magnitudes between 14.5 and 15. We find that the highest redshift of the 346 corresponding sources is $z$=0.9045 and that 93\% of sources have $z<$0.1. Thus, $z$=1 is likely to be far greater than the actual redshift, and 44 pc a conservative overestimate of the linear size.


\begin{table*}[!ht]
\caption{Component properties including spectral characteristics, size estimates from 4.96 GHz images, and magnetic field strengths (calculated with $z$ set to 1 for sources without a known redshift).}
\label{Table:Magnetic_Fields}      
\centering
\newcolumntype{C}{ @{}>{${}}c<{{}$}@{} }
\begin{tabular}{l
                l
                *7{rCl} 
                }
\hline       
\multicolumn{1}{c}{Source,}   & \multicolumn{1}{c}{Proposed}   & \multicolumn{3}{c}{S$_{1.55 \: GHz}$}  & \multicolumn{3}{c}{S$_{4.96 \: GHz}$}  & \multicolumn{3}{c}{${\alpha}_{thin}$} & \multicolumn{3}{c}{$\theta_{D, \: maj}$} & \multicolumn{3}{c}{$\theta_{D, \: min}$} & \multicolumn{3}{c}{$B_{SSA}$} & \multicolumn{3}{c}{$B_{Equi}$}  \\
\multicolumn{1}{c}{Component} & \multicolumn{1}{c}{Morphology} & \multicolumn{3}{c}{(mJy)             } & \multicolumn{3}{c}{(mJy)             } & \multicolumn{3}{c}{                 } & \multicolumn{3}{c}{(mas)    } & \multicolumn{3}{c}{(mas)    } & \multicolumn{3}{c}{(mG)     } & \multicolumn{3}{c}{(mG)      }\\
\hline 
J0231$-0433$, W & \multirow{2}{*}{CSO}      & 150 & \! \pm \! \! & 20 & 37  & \! \pm \! \! & 4   & $-1.2 $ & \! \pm \! \! & 0.2  & 9.0 & \! \pm \! \! & 0.9 & 3.8 & \! \pm \! \! & 0.4 &  0.5   & \! \pm \! \! & 0.4               &  3  & \! \pm \! \! & 3 \\
J0231$-0433$, E &                           & 17  & \! \pm \! \! & 2  & 6.5 & \! \pm \! \! & 0.6 & $-0.8 $ & \! \pm \! \! & 0.1  & 7.6 & \! \pm \! \! & 0.8 & 3.4 & \! \pm \! \! & 0.3 &  0.3   & \! \pm \! \! & 0.2               &  7  & \! \pm \! \! & 4 \\
J0235$-0100$, W & \multirow{2}{*}{CSO}      & 110 & \! \pm \! \! & 10 & 72  & \! \pm \! \! & 7   & $-0.39$ & \! \pm \! \! & 0.06 & 11  & \! \pm \! \! & 1   & 5.4 & \! \pm \! \! & 0.5 &  2     & \! \pm \! \! & 3                 &  9  & \! \pm \! \! & 4 \\
J0235$-0100$, E &                           & 91  & \! \pm \! \! & 9  & 56  & \! \pm \! \! & 6   & $-0.41$ & \! \pm \! \! & 0.06 & 8.6 & \! \pm \! \! & 0.9 & 5.6 & \! \pm \! \! & 0.6 &  2     & \! \pm \! \! & 4                 &  11 & \! \pm \! \! & 5 \\
J0330$-0740$, U &                           & 270 & \! \pm \! \! & 30 & 220 & \! \pm \! \! & 20  & $-0.23$ & \! \pm \! \! & 0.07 & 10  & \! \pm \! \! & 1   & 2.9 & \! \pm \! \! & 0.3 & <0.08  & \! \pm \! \! & 0.04              & >11 & \! \pm \! \! & 5 \\ 
J1509+1406*, W  & \multirow{2}{*}{Core-Jet} & 68  & \! \pm \! \! & 7  & 39  & \! \pm \! \! & 4   & $-0.45$ & \! \pm \! \! & 0.06 & 25  & \! \pm \! \! & 2   & 11  & \! \pm \! \! &   1 &  0.005 & \! \pm \! \! & 0.002             &  8  & \! \pm \! \! & 4 \\ 
J1509+1406*, E  &                           & 97  & \! \pm \! \! & 10 & 26  & \! \pm \! \! & 3   & $-1.1 $ & \! \pm \! \! & 0.2  & 32  & \! \pm \! \! & 3   & 29  & \! \pm \! \! &   3 &  0.07  & \! \pm \! \! & 0.03              &  1  & \! \pm \! \! & 1 \\ 
J1525+0308*, U  &                           & 900 & \! \pm \! \! & 90 & 510 & \! \pm \! \! & 50  & $-0.4 $ & \! \pm \! \! & 0.1  & 11  & \! \pm \! \! & 1   & 5.1 & \! \pm \! \! & 0.5 & <(3    & \! \pm \! \! & 4)$\times10^{-5}$ & >20 & \! \pm \! \! & 10 \\ 
J1548+1320*, U  &                           & 54  & \! \pm \! \! & 5  & 7.6 & \! \pm \! \! & 0.8 & $-0.79$ & \! \pm \! \! & 0.09 & 11  & \! \pm \! \! & 1   & 7.6 & \! \pm \! \! & 0.8 & <0.01  & \! \pm \! \! & 0.03              & >4  & \! \pm \! \! & 2 \\ 
\hline                  
\end{tabular}
\tablefoot{Column 1: source name (J2000, * indicates unknown redshift) and component (U$\equiv$ no clearly resolved second component). 2: proposed morphological classification. 3: integrated flux density at 1.55 GHz. 4: integrated flux density at 4.96 GHz. 5: optically thin spectral index, measured between 1.55 and 4.96 GHz for resolved components. 6: effective deconvolved major axis of the component, with error approximated as the same relative error from Section~\ref{Sec:Results}. 7: effective deconvolved minor axis. 8: magnetic field assuming SSA calculated from Equation~\ref{Equation:B_SSA} (an overestimate for unresolved components). 9: magnetic field assuming equipartition conditions calculated from Equation~\ref{Equation:B_Equi} (an underestimate for unresolved components). For components of resolved sources $B_{SSA}$ has been calculated assuming $S_{peak}$ is approximately the total flux density at $\nu_{peak}$. If we instead assume the power-law derived from 1.55 and 4.96 GHz VLBA flux densities holds at $\nu_{peak}$, $B_{SSA}$ for the Western and Eastern components of J0231-0433, J0235-0100, and J1509+1406 would be 0.04, 7, 4, 6, 0.9, and 0.2 mG, respectively, which still underestimate $B_{Equi}$.}
\end{table*} 


\subsection{J1548+1320}
While J1548+1320 (GLEAM J154852+132058) was resolved, 59$\pm$4\% missing total integrated flux density compared to the curve fit at 1.55 GHz and 85$\pm$2\% at 4.96 GHz suggests that the source was dominated by diffuse emission that has been resolved out. Despite the mottled structure, we include the source for completeness and treat it as one component broadly defined by a 6.4 $\times$ 4.9 mas region. We estimate an upper limit of the linear size to be $\ell {\lesssim}$52 pc assuming $z$=1. It has no infrared or optical counterparts.


\section{Discussion} \label{Sec:Discsussion}
\subsection{Magnetic Field}

Assuming SSA to be the sole mechanism responsible for the spectral turnover, the strength of the magnetic field $B_{SSA}$ can be approximated via
    \begin{equation}
      B_{SSA} \approx \frac{(\nu_{peak} / f(\alpha_{thin}))^{5} \: {\theta_{D, \: maj}}^2 \: {\theta_{D, \: min}}^2}{{S_{peak}}^{2}(1+z)},\label{Equation:B_SSA}
    \end{equation}
\noindent where $B_{SSA}$ is in G, $\nu_{peak}$ is the observed frequency of spectral turnover in GHz, $f(\alpha_{thin})$ is estimated from tabulated values in~\cite{1983ApJ...264..296M} and ranges from ${\sim}$7.7 to ${\sim}$8.9 for our sources, $\theta_{D, \: maj}$ and $\theta_{D, \: min}$ are in mas, and $S_{peak}$ is the flux density at $\nu_{peak}$ in Jy~\citep{1981ARA&A..19..373K}.

Investigating the validity of modeling peaked-spectrum sources using the assumption of minimum energy content, ~\cite{2008A&A...487..885O} found that magnetic fields estimates based on equipartition between particle and magnetic field total energy densities agreed well with estimates based on pure SSA, except in cases were spectra where better modeled by FFA. Such an equipartition magnetic field strength $B_{Equi}$ can be estimated by
    \begin{equation}
      B_{Equi} \approx {\left( \frac{4\pi (2|\alpha_{thin}|+1) (K+1) I_{\nu} {E_p}^{1-2|\alpha_{thin}|}}{ (2|\alpha_{thin}|-1) \: c_2 \: \theta_{D, \: maj} \: c_4 \: (2c_1/\nu)^{|\alpha_{thin}|} } \right),}^{1/(|\alpha_{thin}|+3)}\label{Equation:B_Equi}
    \end{equation}
\noindent where where $B_{Equi}$ is in G, K$\sim$100 (representing the proton-to-electron number density ratio), $E_p$=1.5033$\times10^{-3}$ erg, $\theta_{D, \: maj}$ is in cm, I$_{\nu}$ = S$_{\nu}$/$\Omega$  in erg s$^{-1}$ cm$^{-2}$ Hz$^{-1}$ sr$^{-1}$,  $\Omega{\sim}\frac{\pi \theta_{D, \: maj} \theta_{D, \: min}}{4ln(2)}$ in sr, $c_2$ is a function of $\gamma_e=2|\alpha_{thin}|+1$ and $c_3$, and $c_1$, $c_3$, and $c_4$ are constants (listed in~\citealt{2005AN....326..414B}).

In Table~\ref{Table:Magnetic_Fields}, we compare magnetic field strengths calculated from Equation~\ref{Equation:B_SSA} to Equation~\ref{Equation:B_Equi}. Errors in $B_{SSA}$ and $B_{Equi}$ are estimated through standard propagation of uncertainty, where we have neglected the contribution of constants raised to the power of $|\alpha_{thin}|$ and the error of $\Gamma(x)$ in calculation of $c_2$. Notably, it is important to study the strengths of these fields for each homogeneous component rather than for entire sources~\citep{2008A&A...487..885O}, so in cases where sources are not clearly resolved into more than one component (indicated by `U'), Equation~\ref{Equation:B_SSA} is taken as an overestimate of $B_{SSA}$ based on the strong angular size dependence~\citep{2014MNRAS.438..463O} and Equation~\ref{Equation:B_Equi} is an underestimate given the inverse dependence. Estimates of $B_{Equi}$ from Table~\ref{Table:Magnetic_Fields} take on values typical to those found from $B_{SSA}$ and $B_{Equi}$ estimates for GPS and HFP sources~\citep{2008A&A...487..885O, 2014MNRAS.438..463O}, ${\sim}$1-100 mG. However, we find that $B_{SSA}$ differs from $B_{Equi}$ in every case by factors of $\gtrsim$5 suggesting that Equation~\ref{Equation:B_SSA} does not well describe the actual magnetic fields of our sources. This provides further evidence that SSA does not dominate spectral characteristics below the peak frequency and is not the cause of spectral turnover for our sources.


\subsection{Linear Size} \label{SubSec:LS}
A relationship between linear size and rest-frame turnover frequency has been found for peaked-spectrum sources. By fitting to CSS and GPS sources over 3 orders of magnitude in frequency space~\cite{1997AJ....113..148O} determined the following inverse power-law: 
\begin{equation}
\nu_{rest \: frame \: peak} \approx {10}^{-0.21\pm0.05} \times {\ell}^{-0.65\pm0.05}, \label{Equation:Linear_Size_ODea}
\end{equation}
\noindent where $\nu_{rest \: frame \: peak}=(1+z) \: \nu_{peak}$ is the rest-frame turnover frequency in GHz and $\ell$ is the projected linear size in kpc.~\cite{2014MNRAS.438..463O} derived a similar relationship with $\nu_{rest \: frame \: peak}\propto{{\ell}^{-0.59\pm0.05}}$ for a sample of HFP, GPS, and CSS sources with unambiguous core components. SSA provides for such a relationship since the spectral peak is dependent upon the magnetic field strength which is related to hotspot radius, and therefore linear size, by an inverse power-law~\citep{1997AJ....113..148O}. While FFA can achieve a similar power-law through a model by~\cite{1997ApJ...485..112B}, to exactly match the ${\ell}^{-0.65}$ dependence found by~\cite{1997AJ....113..148O} an implausible medium which increases in density with distance from the galactic nucleus is required (see Subsection~\ref{SubSec:SITM}).

\begin{figure} \includegraphics[width=1.\linewidth]{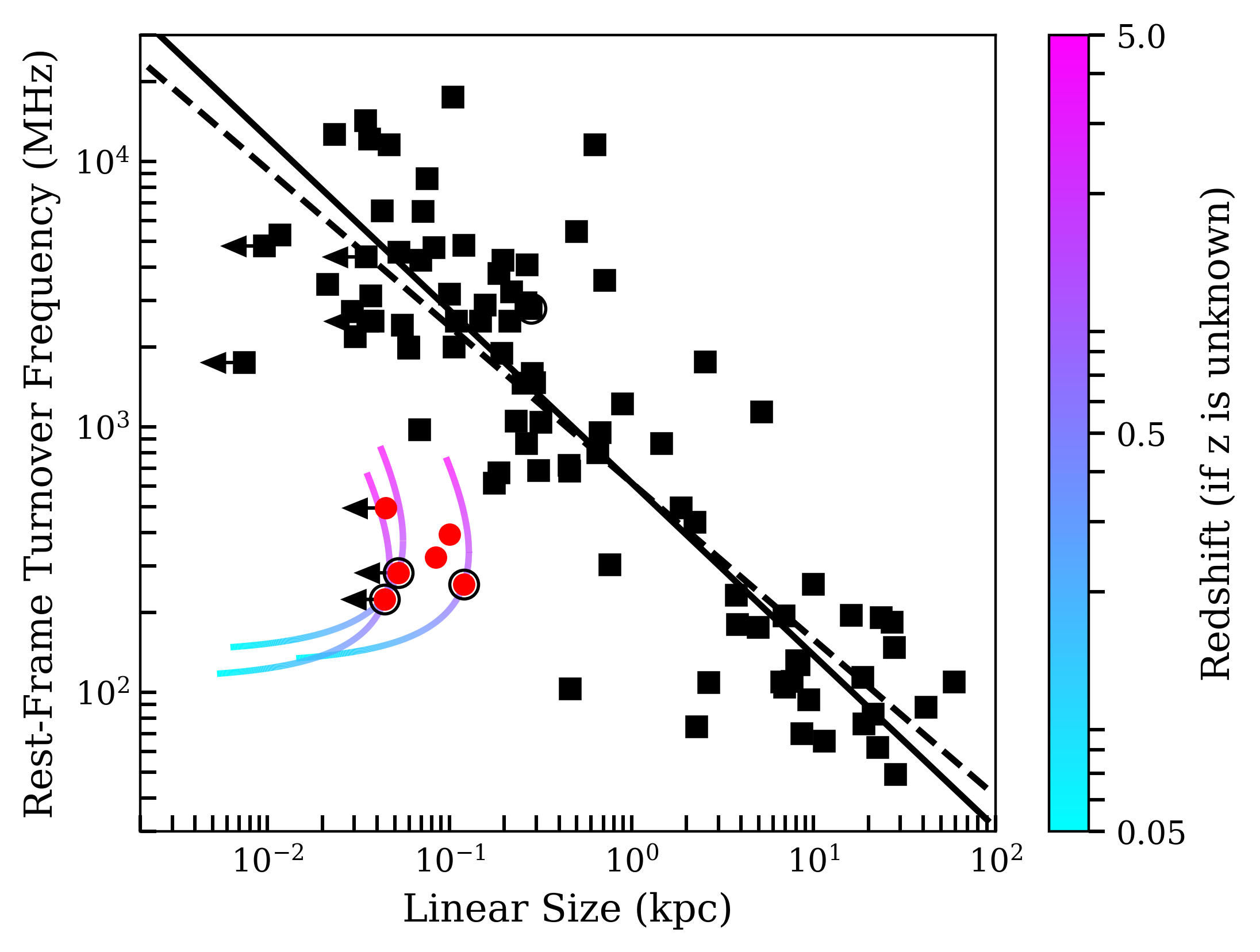}\caption{Rest-frame turnover frequency versus linear size for our sources and those described in Appendix~\ref{Sec:Appendix}. Red circles represent our sources while black squares represent sources from prior literature listed in Table~\ref{Table:Samples}. The solid line represents Equation~\ref{Equation:Linear_Size_ODea}, the dashed line represents the similar relation found by~\cite{2014MNRAS.438..463O}, arrows indicate maximum linear sizes for unresolved sources, and black circles indicate sources with unknown redshifts (set to 1 for this plot). For sources with unknown redshifts, colored lines are included to demonstrate the redshift dependence, with $z$ specified by the color-bar.}\label{Fig:Linear_Size}\end{figure}

In Figure~\ref{Fig:Linear_Size}, we show how our 6 sources compare to the sample from~\cite{1997AJ....113..148O} and additional sources from~\cite{2000MNRAS.319..445S}, with literature values given in Appendix~\ref{Sec:Appendix}. Taking into account the dependence of $z$ in both size projection and frequency shift, the colored lines indicate the redshift dependence for sources with unknown $z$. We find that each of our sources are too small for what Equation~\ref{Equation:Linear_Size_ODea} would suggest based on their rest-frame turnover by factors of 32, 20, >31, 32, >108, and >64 (for J0231-0433, J0235-0100, J0330-0740, J1509+1406, J1525+0308, and J1548+1320, respectively). It is particularly surprising to find that sources with peaks between 72 and 230 MHz, lacking sufficient low frequency coverage to demonstrate SSA-violating thick spectral indices $\alpha_{thick}{\gtrsim}$2.5 reliably, still depart from the linear size, rest-frame turnover frequency anti-correlation. They too appear to be far too small for their rest-frame spectral peak frequency, even at a redshift of $z$ = 5. The most plausible explanation is that the generic curve fit well characterizes the spectra despite the comparatively sparse coverage at below the spectral peak, so all of our sources should be understood as likely SSA violating.

Each of our six sources had linear sizes $\gtrsim$20 times too small than what would be expected based on turnover frequency from previous fits which assumed SSA. This suggests the physics constraining the size of our sources, or causing a MHz spectral turnover, differs from peaked-spectrum sources previously identified at GHz frequencies.

\begin{figure}\centering \includegraphics[width=0.843775809727895711957856202147369360923234893636018661130\linewidth]{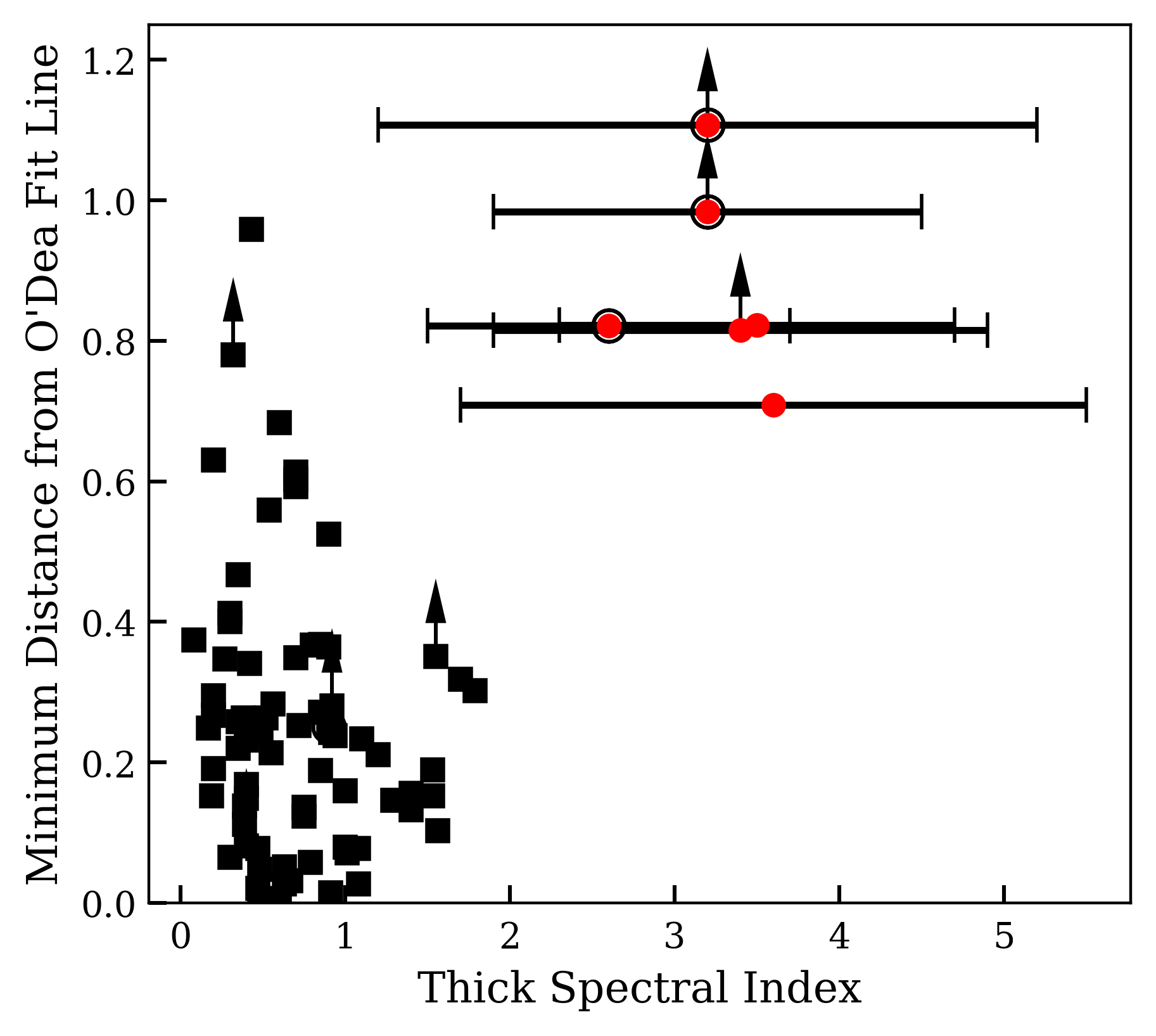}\caption{Minimum distance to Equation~\ref{Equation:Linear_Size_ODea} for each source as calculated in log space (see Equation~\ref{Equation:dmin}). Red circles represent our sources while black squares represent literature sources listed in Table~\ref{Table:Samples}. Arrows indicate lower limits for unresolved sources and black circles indicate sources with unknown redshifts (set to 1 for this plot). Note that the thick spectral indices of our sources with spectral peaks between 72 and 230 MHz are less reliable since their spectra were only sampled at $<$12 frequencies below the spectral turnover.}\label{Fig:dmin}\end{figure}

    
\subsection{Spectral Index and Turnover Mechanism} \label{SubSec:SITM}
We have shown that our sources, with $\alpha_{thick}{\gtrsim}2.5$, violate the empirical linear size, rest-frame turnover frequency anti-correlation predicted by SSA. To test whether such a violation occurs is a function of spectral index, in Figure~\ref{Fig:dmin} we plot sources with known $\alpha_{thick}$ (which excludes all Appendix~\ref{Sec:Appendix} sources with $\nu_{peak}$<250 MHz -- although, since many of the Appendix~\ref{Sec:Appendix} sources were sampled at only 1 frequency below spectral turnover, the accuracy of reported  $\alpha_{thick}$ values are not clear) and show that $\alpha_{thick}$ is not a good indicator of minimum distance $d_{min}$ from the fit line found by~\cite{1997AJ....113..148O}:
\begin{equation}
    d_{min} = \frac{|0.65log_{10}(\ell) + log_{10}(\nu) + 0.21|}{\sqrt{0.65^2 + 1}}. \label{Equation:dmin}
\end{equation}
However, we do find that all our sources significantly violate the linear size, turnover frequency relationship found by~\cite{1997AJ....113..148O}. This may be an indication of a change in the underlying physics of the spectral turnover between the maximum $\alpha_{thick}$=1.8 from Appendix~\ref{Sec:Appendix} and our lowest $\alpha_{thick}$=2.6.

\begin{figure}\centering \includegraphics[width=0.85766423357\linewidth]{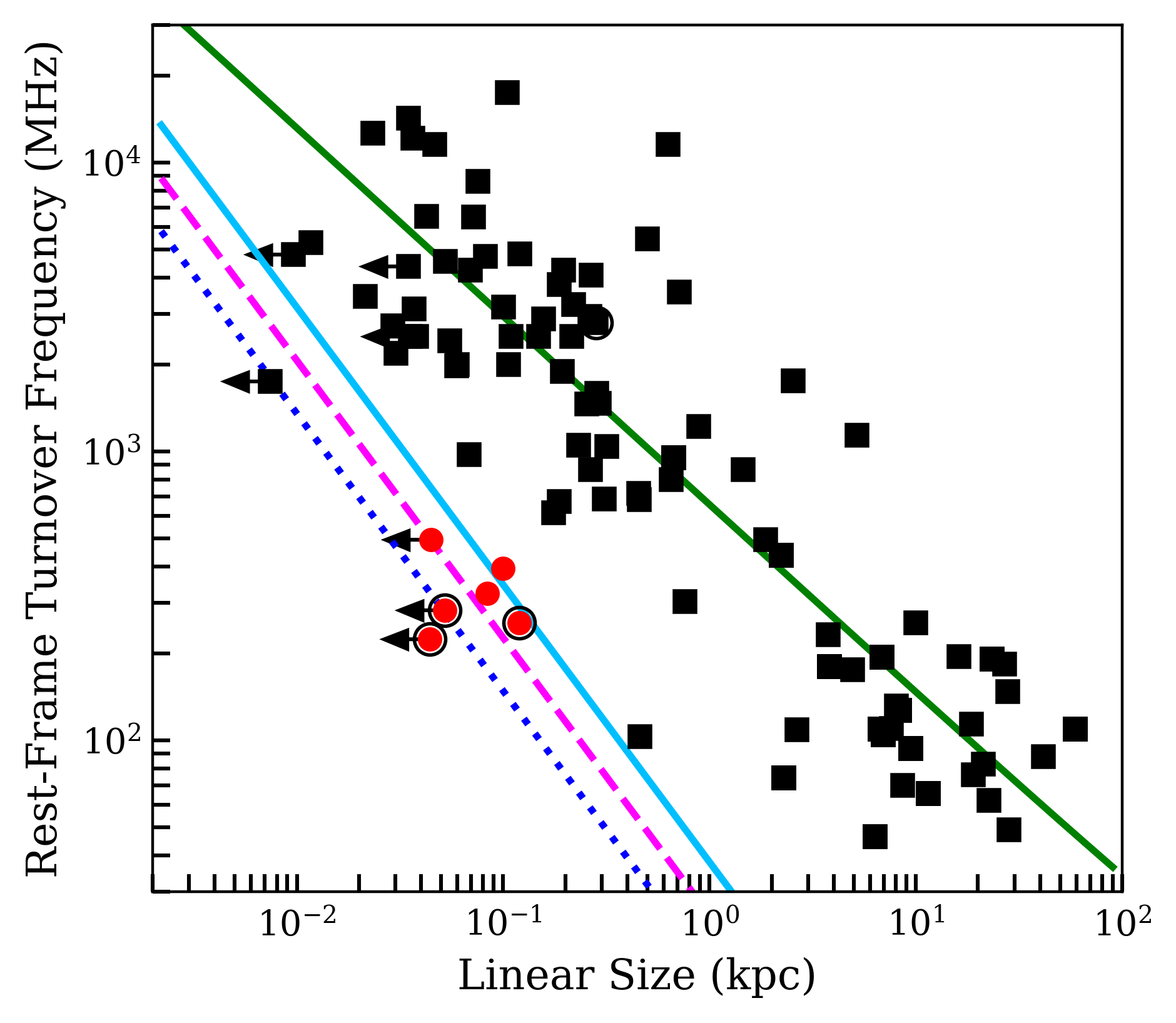}\caption{The turnover frequency, linear size relationship for both our sources and those described in Table~\ref{Table:Samples} with different FFA model parameters. The light blue solid line, the magenta dashed line, and the dark blue dotted line represent FFA models where $E{=}$10$^{43}$, 10$^{42.5}$, and 10$^{42}$ ergs s$^{-1}$, respectively, $\delta{=}$2, and $n_0$ reaches 0.1 cm$^{-3}$ at 1 kpc, which well describe our sources. The green line represents the required FFA model from~\cite{1997ApJ...485..112B} to replicate the power-law slope from Equation~\ref{Equation:Linear_Size_ODea}, where $\delta{=}$-0.75, $n_0{=}$100 cm$^{-3}$ and $E{=}$10$^{45.5}$ ergs s$^{-1}$. Red circles represent our sources, black squares represent sources listed in Table~\ref{Table:Samples}, arrows indicate maximum linear sizes for unresolved sources, and black circles indicate sources with unknown redshifts (set to 1 for this plot).}\label{Fig:Jet}\end{figure}

Given that SSA is unlikely to be operating in our sources based on spectra and magnetic field, FFA is the most likely alternative mechanism~\citep{2015ApJ...809..168C}.~\cite{1997ApJ...485..112B} examines a model in which GPS and CSS proprieties are explained as a consequence of the interaction of radio lobes with the interstellar medium which is photoionized by a bow shock. Under this model, assuming a galactic medium where density decreases with distance $x$ from the source as $x^{-\delta}$ and where the mean lobe pressure, averaged over the hotspot region of the lobe, is $\zeta{\approx}$2 times the average lobe pressure (accounting for the speed of advancement and cross-sectional area of the head of the jet following~\citealt{1996cyga.book..209B}), the expansion velocity $V$ as a function of $x$ can be approximated as
    \begin{equation}
      V \approx  1500 \left( \frac{6}{8-\delta} \right)^{1/3} \zeta^{1/6} \left( \frac{E}{n_0} \right)^{1/3} \left( \frac{x}{x_0} \right)^{(\delta-2)/3}, \label{Equation:V}
    \end{equation}
\noindent where $V$ is in km/s, $x$ is in kpc, $\delta$ is the density profile index, $n_0$ is the Hydrogen number density in cm$^{-3}$ for the shock and precursor regions at $x_0$ = 1 kpc, and $E$ is the jet energy flux in ergs s$^{-1}$. Under the same model, assuming steady state, one-dimensional shocks viewed at normal incidence, the rest frame turnover frequency can be approximated as
    \begin{equation}
      \nu_{rest \: frame \: peak} \approx 1.1 \left( \left( \frac{p+2}{p+1} \right) (aV^{2.3} + bV^{1.5})n_0\left( \frac{x}{x_0} \right)^{-\delta} \right)^{0.48}, \label{Equation:vp}
    \end{equation}
\noindent where $\nu_{rest \: frame \: peak}$ is in GHz, $p{\sim}$-0.17 (representing the power-law distribution of absorbing clouds), and $a$ and $b$ are constants (0.0019 and 0.000997, respectively). $\nu_{rest \: frame \: peak}$ in Equation~\ref{Equation:vp} displays a inverse power-law dependence on $\ell$ and~\cite{1997ApJ...485..112B} found it to qualitatively describe the relationship when $n_0{\approx}$10-100, $log_{10}(E){\approx}$45-46 and $\delta{\approx}$1.5-2. However, the model requires a $\delta$<-0.7, for which the medium would \textit{increase} in density with distance from the nucleus, to exactly replicate the ${\ell}^{-0.65}$ dependence found by~\cite{1997AJ....113..148O}. In Figure~\ref{Fig:Jet}, the green line represents a model with $\delta{=}$-0.75, $n_0{=}$100 cm$^{-3}$ and $E{=}$10$^{45.5}$ ergs s$^{-1}$, which well approximates the linear size-turnover relationship.

While the FFA model examined by~\cite{1997ApJ...485..112B} may not perfectly describe the relationship found by~\cite{1997AJ....113..148O} it still provides insight into medium properties for our sources which do not have spectra or magnetic fields well described by SSA. Since the spectral turnover estimated from Equation~\ref{Equation:vp} is positively dependent on medium density due to the free-free optical depth's dependence on $n_0$ of both the shock region and the jet precursor region, we find that, in order to replicate the MHz rest-frame turnover frequencies for our sources, a medium that reaches $n_0{=}$0.1 cm$^{-3}$ at 1 kpc is sufficient. In Figure~\ref{Fig:Jet}, we demonstrate how such a model of FFA with $\delta{=}$2 and `weaker' jet energies $E{=}$10$^{42}$, 10$^{42.5}$, and 10$^{43}$ ergs s$^{-1}$ (similar to those explored by~\citealt{2007ApJS..173...37S} and~\citealt{2011ApJ...728...29W}) can explain the turnover frequencies and linear sizes for our sources.


\subsection{Luminosity and the Evolutionary Model}
In addition to framing turnover frequency as a function of source size, and therefore age, proposed evolutionary models have also considered the variation of luminosity as function of size~\citep{1997AJ....113..148O, 2010MNRAS.408.2261K}. Under the model presented by~\cite{2012ApJ...760...77A}, the youngest HFP and GPS sources would increase luminosity with age due to an increasingly more efficient transformation of jet kinetic energy into radiative emission. Once the sources grow into CSS sources, they would eventually reach a balance between adiabatic losses and synchrotron losses, and their luminosity would become constant in time. Finally, once sources become characteristic of FR I or II AGN, their luminosity would decrease in time as the inter-galactic medium at the front of the jet becomes increasingly less dense.~\cite{2010MNRAS.408.2261K} further explored the CSS population and presented a model in which high luminosity CSS sources grow into FR II AGN while short-lived, low luminosity CSS sources become FR I AGN.

\begin{figure} \centering \includegraphics[width=0.847171532840154255319148936170212765957446808510638297872\linewidth]{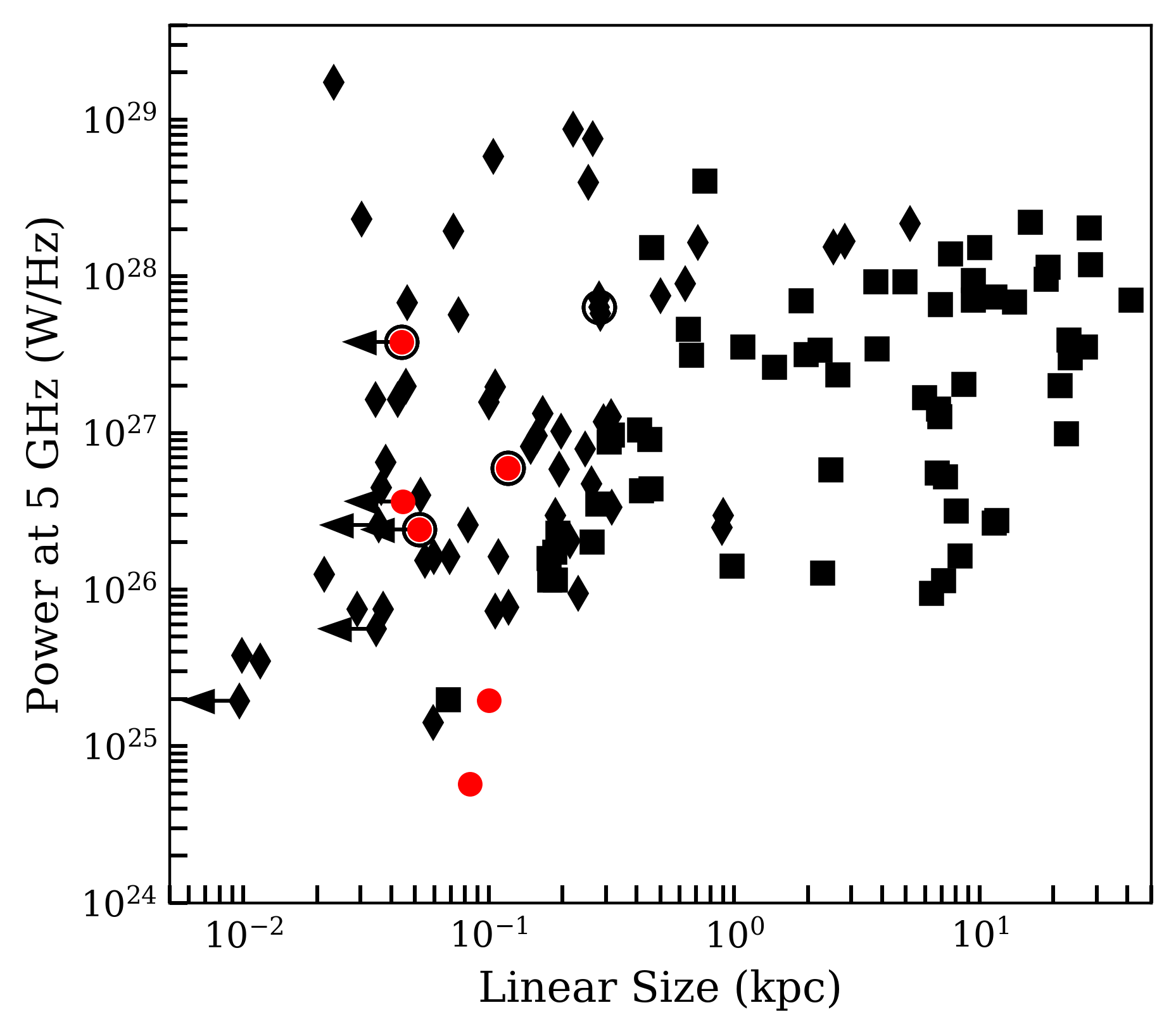}\caption{Luminosity versus size for both our sources and those described in Table~\ref{Table:Samples}. Red circles represent our sources. Black diamonds and squares represent sources listed in Table~\ref{Table:Samples} which peak above (GPS) or below (CSS) 1 GHz, respectively. Arrows indicate maximum linear sizes for unresolved sources and black circles indicate sources with unknown redshifts (set to 1 for this plot).}\label{Fig:LP}\end{figure}

In Figure~\ref{Fig:LP} we show how our sources fit into the GPS and (high luminosity) CSS populations based on linear size and radio power, distinguishing CSS sources as those which peak below 1 GHz. Incorporating standard adjustment based on luminosity distance and $k$-correction, we calculate radio power $P_{5 \: GHz}$ as 
\begin{equation}
    P_{5 \: GHz} = 4 \pi D_L^2 S_{5 \: GHz} (1+z)^{-(1+\alpha_{thin})}
\end{equation}
where $P_{5 \: GHz}$ is in W Hz$^{-1}$, $D_L$ is the luminosity distance in m, and $S_{5 \: GHz}$ is the flux density at 5 GHz in W m$^{-2}$ Hz$^{-1}$, calculated for our sources from the generic curve model due to the missing flux densities reported in Section~\ref{Sec:Results}.

Similar to our findings in Subsection~\ref{SubSec:LS}, we find our sources fit well into the trend seen in GPS sources, and departs from the CSS population, even though our sources should fit into the CSS population based on turnover frequency. Interestingly, the low luminosities of our two resolved sources with reliably SSA-violating optically thick spectral indices suggest they might evolve under the low luminosity paradigm proposed by~\cite{2010MNRAS.408.2261K}, undergoing disturbed evolution and starting to fade away as single-lobed objects or becoming FR I AGN or hybrids. However, since the sources already violate the model based on their size and turnover frequency, it is not clear if this evolutionary path could still be expected.

\section{Conclusions} \label{Sec:Conclusions}
We selected six MPS sources from the GLEAM survey to study how they fit within the general radio galaxy evolution picture. Three sources had spectral characteristics which violate SSA and three that had spectral peaks below 230 MHz. After VLBI imaging of VLBA observations at 1.55 and 4.96 GHz, we resolved two of the SSA-violating sources and found they are likely CSOs, while the only resolved source with a peak below 230 MHz is likely a core-jet. The main results of this study are as follows:
\begin{itemize}
\item The sources were all at least one order of magnitude too small compared to the linear size, turnover frequency anti-correlation found for GPS and CSS samples. They present as unique outliers compared to the previously explored peaked-spectrum population. In particular, this suggests that the sources with spectral peaks below 230 MHz should also be understood as SSA-violating despite poor low frequency spectral coverage. 
\item For each of our sources, whereas equipartition approximations of magnetic field strengths were comparable to those of previously studied peaked-spectrum sources, SSA predicts values differing in some cases by several orders of magnitude. Such a discrepancy further demonstrates that SSA is likely not the cause of turnover for our sources.
\item The sources all fit within the GPS, rather than CSS, population based on linear size and luminosity despite their megahertz spectral peaks. 
\item The turnover frequencies and linear sizes of all six sources could be explained under a simplistic model of external FFA by an ionized medium.
\end{itemize}
These findings suggest that FFA is the most likely absorption mechanism responsible for the spectral peak of our sources. It is therefore possible they are small due to interaction with the surrounding medium rather than youth. A population of similar sources would help explain the excess of peaked-spectrum sources compared to larger radio galaxies. Observations of a larger sample of MPS sources with gigahertz-sensitive VLBI arrays, and at low-frequencies with such telescopes as LOFAR, will be necessary to assess whether our target sources are reflective of the MPS population as a whole.


\begin{acknowledgements}
       We thank the Joint Institute for VLBI European Research Infrastructure Consortium staff, Jay Blanchard in particular, for guidance in synthesis imaging. 
       
        M.~A.~Keim thanks Leiden Observatory and ASTRON for their support of travel costs.
       
       We thank Dale Frail and summer students at the National Radio Astronomy Observatory for organizing and conducting the VLBA observations described in Section~\ref{Sec:Observations}. 
       
       The VLBA is an instrument of the National Radio Astronomy Observatory. The National Radio Astronomy Observatory is a facility of the National Science Foundation operated by Associated Universities, Inc. 
       
       This research has made use of the NASA/IPAC Extragalactic Database and Infrared Science Archive which are operated by the Jet Propulsion Laboratory, California Institute of Technology, under contract with the National Aeronautics and Space Administration, the VizieR catalog access tool, CDS, Strasbourg, France, Topcat~\citep{2005ASPC..347...29T, 2006ASPC..351..666T}, CosmoCalc ~\citep[translated into Python by J. Schombert]{2006PASP..118.1711W}, and Astropy, a community-developed core Python package for Astronomy~\citep{2013A&A...558A..33A, 2018AJ....156..123A}. 
       
       This publication makes use of data products from the Wide-field Infrared Survey Explorer, which is a joint project of the University of California, Los Angeles, and the Jet Propulsion Laboratory/California Institute of Technology, funded by the National Aeronautics and Space Administration and from the Two Micron All Sky Survey, which is a joint project of the University of Massachusetts and the Infrared Processing and Analysis Center/California Institute of Technology, funded by the National Aeronautics and Space Administration and the National Science Foundation.
\end{acknowledgements}


\bibliographystyle{aa}
\bibliography{AA-2019-36107.bbl}

\clearpage
\onecolumn
\begin{appendix}
\section{Peaked-spectrum source samples} \label{Sec:Appendix}
\LTcapwidth=0.91\textwidth
\begin{longtable}{l l l l l l l l l l l}
\caption{Samples used for comparison to the anti-correlation between linear size and rest-frame turnover frequency in Section~\ref{Sec:Discsussion}. Original samples have been modified to reflect the cosmological model described in Section~\ref{Sec:Introduction}. Column 1: source name in reported Besselian epoch and Julian conversion (where B1950 declination has 3 significant figures, J2000 conversion is only accurate to the nearest 10$^{1}$ arcminute). 2: references (A $\equiv$~\citealt{1997AJ....113..148O}, B $\equiv$~\citealt{2000MNRAS.319..445S}, C $\equiv$~\citealt{1998A&AS..131..435S}, D $\equiv$~\citealt{1997A&A...321..105D}, E $\equiv$~\citealt{1998A&AS..131..303S}, or F $\equiv$~\citealt{1990A&A...231..333F}) and optical identification (Q $\equiv$ Quasar, G $\equiv$ Galaxy, and U $\equiv$ Unlisted in reference). 3: redshift (set to 1 if unknown). 4: angular size. 5: observed frequency of spectral peak. 6: projected linear size. 7: turnover flux density. 8: thick spectral index. 9: thin spectral index. 10: power at 5 GHz.}\label{Table:Samples} \\       
\hline
\multicolumn{2}{c}{Source}                                  & \multicolumn{1}{c}{Refs, ID} & \multicolumn{1}{c}{$z$} & \multicolumn{1}{c}{$\theta$} & \multicolumn{1}{c}{${\nu}_{peak}$} & \multicolumn{1}{c}{$\ell$} & \multicolumn{1}{c}{S$_{peak}$} & \multicolumn{1}{c}{${\alpha}_{thick}$} & \multicolumn{1}{c}{${\alpha}_{thin}$} & \multicolumn{1}{c}{$P_{5 \: GHz}$} \\
\multicolumn{1}{c}{(B1950)} & \multicolumn{1}{c}{(J2000)}   &                           &                       & \multicolumn{1}{c}{(mas)}    & \multicolumn{1}{c}{(MHz)}          & \multicolumn{1}{c}{(kpc)}       & \multicolumn{1}{c}{(Jy)}       &                                        & & \multicolumn{1}{c}{(W/Hz)} \\
\hline
B0019$-000$ & J0022+0017 & B \& E, G & 0.305 & 70.0 & 800 & 0.317 & 3.47 & 1.0 & $-1.2$ & 3.3$\times10^{26}$ \\
B0108+388 & J0111+3904 & B \& D, G & 0.669 & 6.0 & 3900 & 0.0427 & 1.3 & 0.39 & $-0.54$ & 1.6$\times10^{27}$ \\
B0127+233 & J0130+2334 & A \& F, Q & 1.459 & 2600 & 20.0 & 28.3 & ... & ... & $-0.71$ & 1.2$\times10^{28}$ \\
B0134+329 & J0137+3309 & A \& F, Q & 0.367 & 500 & 80.0 & 2.65 & ... & ... & $-0.82$ & 2.3$\times10^{27}$ \\
B0138+138 & J0141+1403 & B \& F, G & 0.62 & 1000 & 120 & 6.87 & ... & ... & $-0.89$ & 1.3$\times10^{27}$ \\
B0221+276 & J0224+2750 & B \& F, G & 0.31 & 2500 & 50.0 & 11.5 & ... & ... & $-0.93$ & 2.7$\times10^{26}$ \\
B0237$-233$ & J0239$-2229$ & A \& D, Q & 2.223 & 18.0 & 1000 & 0.22 & 7.0 & 0.49 & $-0.64$ & 8.7$\times10^{28}$ \\
B0248+430 & J0251+4312 & A \& D, Q & 1.316 & 60.0 & 5000 & 0.63 & 1.5 & 0.43 & $-0.53$ & 9.0$\times10^{27}$ \\
B0316+161 & J0319+1617 & B \& E, G & 1.2 & 300 & 800 & 2.54 & 9.55 & 0.7 & $-1.1$ & 1.5$\times10^{28}$ \\
B0319+121 & J0322+1217 & A \& F, Q & 2.662 & 20.0 & 400 & 0.254 & ... & ... & $-0.59$ & 4.0$\times10^{28}$ \\
B0345+337 & J0348+3351 & B \& F, G & 0.24 & 600 & 60.0 & 2.29 & ... & ... & $-0.84$ & 1.3$\times10^{26}$ \\
B0400+6042 & J0404+6050 & B \& C, U & 1.5 & 4.4 & 1000 & 0.0381 & 0.184 & 0.52 & $-1.48$ & 6.5$\times10^{26}$ \\
B0404+76 & J0411+76 & B \& E, G & 0.59 & 100 & 600 & 0.671 & 8.15 & 0.3 & $-0.5$ & 3.1$\times10^{27}$ \\
B0428+205 & J0431+2037 & B \& D, G & 0.219 & 250 & 1000 & 0.889 & 4.0 & 0.35 & $-0.33$ & 2.5$\times10^{26}$ \\
B0429+415 & J0432+4137 & A \& F, Q & 1.023 & 80.0 & 150 & 0.76 & ... & ... & $-2.0$ & 4.1$\times10^{28}$ \\
B0436+6152 & J0441+6158 & B \& C, U & 1.5 & 17.1 & 1000 & 0.148 & 0.237 & 0.79 & $-1.01$ & 8.2$\times10^{26}$ \\
B0500+019 & J0503+0158 & B \& D, G & 0.583 & 15.0 & 2000 & 0.1 & 2.6 & 0.63 & $-0.32$ & 1.6$\times10^{27}$ \\
B0518+165 & J0521+1633 & A \& F, Q & 0.76 & 600 & 100 & 4.95 & ... & ... & $-0.65$ & 9.2$\times10^{27}$ \\
B0535+6743 & J0540+6745 & B \& C, U & 1.5 & 4.0 & 5700 & 0.0346 & 0.192 & 0.27 & $-0.94$ & 1.6$\times10^{27}$ \\
B0538+498 & J0542+4950 & A \& F, Q & 0.545 & 550 & 150 & 3.77 & ... & ... & $-0.83$ & 9.2$\times10^{27}$ \\
B0539+6200 & J0544+6202 & B \& C, U & 1.4 & 6.1 & 1900 & 0.0526 & 0.129 & 0.67 & $-0.33$ & 4.0$\times10^{26}$ \\
B0552+6017 & J0557+6018 & B \& C, U & 1.5 & 12.6 & 1000 & 0.109 & 0.05 & 0.91 & $-1.16$ & 1.6$\times10^{26}$ \\
B0557+5717 & J0601+5717 & B \& C, U & 1.2 & 6.5 & 1100 & 0.055 & 0.069 & 0.85 & $-1.2$ & 1.5$\times10^{26}$ \\
B0710+439 & J0714+4349 & B \& D, G & 0.518 & 25.0 & 1900 & 0.157 & 2.2 & 0.75 & $-0.28$ & 9.6$\times10^{26}$ \\
B0738+313 & J0741+3111 & A \& E, Q & 0.63 & 10.0 & 5300 & 0.0752 & 3.82 & 0.7 & $-0.8$ & 5.7$\times10^{27}$ \\
B0740+380 & J0743+3753 & A \& F, Q & 1.063 & 2200 & 40.0 & 21.2 & ... & ... & $-1.08$ & 2.0$\times10^{27}$ \\
B0743$-006$ & J0746+0029 & A \& D, Q & 0.994 & 5.0 & 5800 & 0.0465 & 2.0 & 0.42 & $-0.3$ & 6.8$\times10^{27}$ \\
B0752+6355 & J0757+6347 & B \& C, U & 0.9 & 4.6 & 6400 & 0.0365 & 0.314 & 1.79 & $-0.1$ & 4.5$\times10^{26}$ \\
B0758+143 & J0801+1410 & A \& F, Q & 1.197 & 4100 & 40.0 & 41.5 & ... & ... & $-0.89$ & 7.0$\times10^{27}$ \\
B0759+6557 & J0804+6549 & B \& C, U & 1.5 & 8.0 & 1700 & 0.0693 & 0.046 & 1.01 & $-0.97$ & 1.6$\times10^{26}$ \\
B0830+5813 & J0834+5803 & B \& C, U & 0.093 & <4.3 & 1600 & <0.00745 & 0.065 & 0.32 & $-0.51$ & 7.1$\times10^{23}$ \\
B0941$-080$ & J0943$-0814$ & B \& E, G & 0.228 & 48.0 & 500 & 0.176 & 3.4 & 0.3 & $-1.0$ & 1.6$\times10^{26}$ \\
B1005+077 & J1008+0727 & B \& F, G & 0.88 & 1200 & 50.0 & 9.44 & ... & ... & $-0.91$ & 7.0$\times10^{27}$ \\
B1019+222 & J1022+2157 & B \& F, G & 1.62 & 800 & 40.0 & 6.94 & ... & ... & $-1.28$ & 6.6$\times10^{27}$ \\
B1031+567 & J1034+5627 & B \& D, G & 0.459 & 33.0 & 1300 & 0.194 & 2.0 & 0.47 & $-0.32$ & 5.8$\times10^{26}$ \\
B1117+146 & J1120+1420 & B \& E, G & 0.362 & 90.0 & 500 & 0.458 & 3.89 & 0.4 & $-0.8$ & 4.4$\times10^{26}$ \\
B1127$-145$ & J1130$-1347$ & A \& E, Q & 1.187 & 3.0 & 1000 & 0.0303 & 5.8 & 0.8 & $-0.6$ & 2.3$\times10^{28}$ \\
B1143$-245$ & J1146$-2347$ & A \& D, Q & 1.95 & 6.0 & 2200 & 0.0717 & 1.6 & 1.1 & $-0.33$ & 1.9$\times10^{28}$ \\
B1153+317 & J1156+3125 & A \& F, Q & 1.557 & 900 & 100 & 9.99 & ... & ... & $-0.92$ & 1.5$\times10^{28}$ \\
B1203+645 & J1206+6413 & B \& F, G & 0.37 & 1300 & 80.0 & 6.71 & ... & ... & $-0.97$ & 5.5$\times10^{26}$ \\
B1225+368 & J1227+3631 & A \& D, Q & 1.974 & 60.0 & 1200 & 0.712 & 2.1 & 0.54 & $-0.7$ & 1.6$\times10^{28}$ \\
B1245$-197$ & J1248$-1834$ & A \& E, Q & 1.28 & 500 & 500 & 5.19 & 8.69 & 0.7 & $-0.9$ & 2.2$\times10^{28}$ \\
B1250+568 & J1252+5632 & A \& F, Q & 0.321 & 1670 & 100 & 8.05 & ... & ... & $-0.72$ & 3.2$\times10^{26}$ \\
B1323+321 & J1325+3150 & B \& D, G & 0.369 & 60.0 & 500 & 0.309 & 7.5 & 0.4 & $-0.5$ & 8.8$\times10^{26}$ \\
B1328+254 & J1330+2508 & A \& F, Q & 1.055 & 48.0 & 50.0 & 0.461 & ... & ... & $-0.63$ & 1.5$\times10^{28}$ \\
B1328+307 & J1330+3026 & A \& F, Q & 0.849 & 3200 & 80.0 & 27.9 & ... & ... & $-0.55$ & 2.0$\times10^{28}$ \\
B1345+125 & J1347+1215 & B \& E, G & 0.122 & 85.0 & 600 & 0.187 & 8.86 & 0.9 & $-0.7$ & 1.2$\times10^{26}$ \\
B1358+624 & J1360+6209 & B \& D, G & 0.431 & 80.0 & 500 & 0.454 & 6.5 & 1.4 & $-0.56$ & 9.1$\times10^{26}$ \\
B1404+286 & J1406+2822 & B \& D, G & 0.077 & 8.0 & 4900 & 0.0117 & 3.0 & 0.85 & $-0.51$ & 3.5$\times10^{25}$ \\
B1416+067 & J1418+0628 & A \& F, Q & 1.439 & 1490 & 80.0 & 16.1 & ... & ... & $-1.1$ & 2.2$\times10^{28}$ \\
B1419+4158 & J1421+4144 & B \& F, G & 0.37 & 11500 & 80.0 & 59.3 & ... & ... & $-0.88$ & 4.4$\times10^{26}$ \\
B1442+101 & J1444+0953 & A \& D, Q & 3.535 & 20.0 & 900 & 0.266 & 2.5 & 0.08 & $-0.56$ & 7.5$\times10^{28}$ \\
B1443+77 & J1442+77 & B \& F, G & 0.27 & 2000 & 100 & 8.32 & ... & ... & $-1.07$ & 1.6$\times10^{26}$ \\
B1447+77 & J1446+77 & B \& F, G & 1.13 & 2800 & 90.0 & 23.5 & ... & ... & $-1.07$ & 3.0$\times10^{27}$ \\
B1458+718 & J1458+7136 & A \& F, Q & 0.905 & 2110 & 40.0 & 19.0 & ... & ... & $-0.6$ & 1.1$\times10^{28}$ \\
B1517+204 & J1519+2013 & A \& F, G & 0.752 & 1050 & 40.0 & 8.62 & ... & ... & $-0.92$ & 2.0$\times10^{27}$ \\
B1525+6801 & J1525+6750 & B \& C, U & 1.1 & 22.4 & 1800 & 0.187 & 0.163 & 0.38 & $-1.07$ & 3.0$\times10^{26}$ \\
B1551+6822 & J1551+6813 & B \& C, U & 1.3 & 2.5 & 1500 & 0.0214 & 0.052 & 0.56 & $-1.65$ & 1.2$\times10^{26}$ \\
B1557+6220 & J1558+6211 & B \& C, U & 0.9 & <4.4 & 2300 & <0.0349 & 0.049 & 0.4 & $-1.89$ & 5.6$\times10^{25}$ \\
B1600+335 & J1602+3322 & B \& E, G & 1.1 & 60.0 & 2600 & 0.5 & 3.06 & 0.2 & $-0.9$ & 7.5$\times10^{27}$ \\
B1600+7131 & J1560+7122 & B \& C, U & 1.5 & 22.7 & 1700 & 0.197 & 0.346 & 1.7 & $-1.56$ & 1.0$\times10^{27}$ \\
B1607+268 & J1609+2640 & B \& D, G & 0.473 & 49.0 & 1000 & 0.293 & 5.0 & 1.08 & $-0.6$ & 1.2$\times10^{27}$ \\
B1620+6406 & J1620+6359 & B \& C, U & 1.2 & 14.2 & 2200 & 0.12 & 0.047 & 0.17 & $-1.56$ & 7.7$\times10^{25}$ \\
B1622+6630 & J1622+6623 & B \& C, U & 0.201 & <2.9 & 4000 & <0.00965 & 0.363 & 1.55 & $-0.46$ & 2.0$\times10^{25}$ \\
B1634+628 & J1635+6242 & A \& F, Q & 0.988 & 200 & 250 & 1.87 & ... & ... & $-0.84$ & 7.0$\times10^{27}$ \\
B1637+626 & J1638+6230 & B \& F, G & 0.75 & 300 & 250 & 2.23 & ... & ... & $-0.99$ & 3.4$\times10^{27}$ \\
B1639+6711 & J1639+6705 & B \& C, U & 1.5 & <4.1 & 1000 & <0.0355 & 0.068 & 0.92 & $-0.61$ & 2.6$\times10^{26}$ \\
B1655+6446 & J1655+6441 & B \& C, U & 1.5 & 24.8 & 1000 & 0.215 & 0.069 & 1.29 & $-0.99$ & 2.0$\times10^{26}$ \\
B1657+5826 & J1658+5821 & B \& C, U & 1.1 & 27.8 & 500 & 0.232 & 0.064 & 0.19 & $-0.56$ & 9.4$\times10^{25}$ \\
B1807+5959 & J1808+5959 & B \& C, U & 1.0 & 13.0 & 1000 & 0.106 & 0.047 & 1.56 & $-0.9$ & 7.3$\times10^{25}$ \\
B1807+6742 & J1807+6742 & B \& C, U & 1.5 & 6.9 & 800 & 0.0598 & 0.054 & 0.94 & $-0.7$ & 1.6$\times10^{26}$ \\
B1808+6813 & J1808+6813 & B \& C, U & 1.1 & 3.5 & 1300 & 0.0292 & 0.042 & 0.2 & $-0.55$ & 7.5$\times10^{25}$ \\
B1819+39 & J1821+39 & B \& F, G & 0.8 & 500 & 100 & 3.81 & ... & ... & $-1.02$ & 3.4$\times10^{27}$ \\
B1819+6707 & J1819+6708 & B \& C, U & 0.22 & 19.2 & 800 & 0.0685 & 0.338 & 0.35 & $-0.54$ & 2.0$\times10^{25}$ \\
B1829+29 & J1831+29 & A \& F, Q & 0.842 & 3100 & 100 & 26.9 & ... & ... & $-0.85$ & 3.5$\times10^{27}$ \\
B1841+6715 & J1841+6718 & B \& C, U & 0.486 & 6.1 & 2100 & 0.037 & 0.178 & 1.53 & $-0.63$ & 7.5$\times10^{25}$ \\
B1843+6305 & J1843+6308 & B \& C, U & 1.5 & 9.5 & 1900 & 0.0823 & 0.075 & 1.53 & $-0.9$ & 2.6$\times10^{26}$ \\
B1942+7214 & J1941+7221 & B \& C, U & 1.1 & 31.5 & 1400 & 0.262 & 0.233 & 0.72 & $-0.5$ & 4.7$\times10^{26}$ \\
B1946+7048 & J1946+7055 & B \& C, U & 0.101 & 31.9 & 1800 & 0.0595 & 0.929 & 0.91 & $-0.64$ & 1.4$\times10^{25}$ \\
B2008$-068$ & J2011$-0503$ & A \& E, G & $\equiv$1 & 30.0 & 1400 & 0.282 & 2.64 & 0.9 & $-0.8$ & 6.3$\times10^{27}$ \\
B2126$-158$ & J2129$-1359$ & A \& D, Q & 3.27 & 8.0 & 4100 & 0.104 & 1.2 & 0.6 & $-0.51$ & 5.8$\times10^{28}$ \\
B2128+048 & J2131+0501 & B \& D, G & 0.99 & 35.0 & 800 & 0.285 & 5.0 & 0.48 & $-0.42$ & 5.8$\times10^{27}$ \\
B2134+004 & J2137+0037 & A \& E, Q & 1.936 & 2.0 & 4300 & 0.0234 & 8.59 & 1.2 & $-0.7$ & 1.7$\times10^{29}$ \\
B2210+016 & J2213+0151 & B \& E, G & 1.0 & 80.0 & 400 & 0.652 & 4.51 & 0.6 & $-1.0$ & 4.6$\times10^{27}$ \\
B2248+71 & J2250+71 & A \& F, G & 1.841 & 1600 & 40.0 & 18.7 & ... & ... & $-1.24$ & 9.6$\times10^{27}$ \\
B2249+185 & J2251+1846 & A \& F, Q & 1.758 & 660 & 40.0 & 7.6 & ... & ... & $-0.76$ & 1.4$\times10^{28}$ \\
B2252+129 & J2254+1310 & A \& F, Q & 0.543 & 3300 & 40.0 & 22.6 & ... & ... & $-1.06$ & 9.8$\times10^{26}$ \\
B2342+821 & J2344+8223 & A \& D, Q & 0.735 & 180 & 500 & 1.46 & 5.0 & 0.55 & $-0.62$ & 2.6$\times10^{27}$ \\
B2352+495 & J2355+4947 & B \& E, G & 0.237 & 70.0 & 700 & 0.264 & 2.93 & 0.2 & $-0.5$ & 2.0$\times10^{26}$ \\
\hline
\end{longtable}
\end{appendix}


\end{document}